\documentclass[lettersize,journal]{IEEEtran}

\usepackage[numbers,sort&compress]{natbib}

\usepackage{algorithmic}
\usepackage{graphicx}
\usepackage{subcaption}
\usepackage{textcomp}
\usepackage{bmpsize}
\usepackage{xcolor}
\usepackage{lipsum}
\usepackage{url}

\usepackage{listings}
\usepackage[most]{tcolorbox}
\usepackage{pifont} % add this in the preamble
\usepackage{multirow}   % for merging rows
\usepackage{booktabs}   % for \toprule, \midrule, \bottomrule
\usepackage{xspace}
\usepackage{makecell}

\usepackage{orcidlink}
\newcommand{\fuzzer}{Agnostic Fuzzer\xspace}
\newcommand{\tracelib}{TraceLib\xspace}

\newcommand{\tracelibplain}{TraceLib\_\allowbreak plain}
\newcommand{\tracelibprojected}{TraceLib\_\allowbreak projected}
\newcommand{\tracelibngram}{TraceLib\_\allowbreak nGram}

\begin{document}
% \let\WriteBookmarks\relax
% \def\floatpagepagefraction{1}
% \def\textpagefraction{.001}

% \shorttitle{TraceLib: Syscall-to-Bitmap Coverage Feedback}
% \shortauthors{I.P.A. Dharmaadi, et~al.}

% \title [mode = title]{TraceLib: Syscall-to-Bitmap Coverage Feedback for Language-Agnostic Web Fuzzing}  
% \title [mode = title]{TraceLib: System-Call Bitmap Feedback Mechanism for Language-Agnostic Web Fuzzing}

\title{TraceLib: System-Call Bitmap Feedback Mechanism for Language-Agnostic Web Fuzzing}

% \author[1,3]{I Putu Arya Dharmaadi}
% \ead{arya.dharmaadi@rug.nl}

% \affiliation[1]{organization={Bernoulli Institute for Mathematics, Computer Science and Artificial Intelligence, University of Groningen},
%                 % city={Groningen},
%                 country={the Netherlands}}

% \author[2]{Elias Athanasopoulos}

% \affiliation[2]{organization={Department of Computer Science, University of Cyprus},
%                 city={Nicosia},
%                 country={Cyprus}}

% \author[1]{Fatih Turkmen}

% \affiliation[3]{organization={Department of Information Technology, Udayana University},
%                 city={Badung},
%                 state={Bali}, 
%                 country={Indonesia}}

%\title{Agnostic Web Fuzzer: Guiding Fuzzing with Syscall-level Tracing}
% \title{Agnostic Web Fuzzer: Guiding Fuzzing with Syscall-level Tracing and SQL Query}
% \title{Example submission for Euro S\&P*\\ {\footnotesize
%     \textsuperscript{*}Note: Sub-titles are not captured in Xplore and
%     should not be used} \thanks{NB: appendices, if any, did
%     not benefit from peer review.}  }

\author{\IEEEauthorblockN{I Putu Arya Dharmaadi\orcidlink{0000-0001-8284-6138}, Elias Athanasopoulos\orcidlink{0000-0002-8759-3261}, and Fatih Turkmen\orcidlink{0000-0002-6262-4869}}
\thanks{First Author is with the Bernoulli Institute, University of Groningen, the Netherlands, and the Faculty of Engineering, Udayana University, Indonesia.
Email: arya.dharmaadi@rug.nl.}
\thanks{Second Author is with the Department of Computer Science, University of Cyprus, Cyprus.}
\thanks{Third Author is with the Bernoulli Institute, University of Groningen, the Netherlands.}
}

% \author{
% \IEEEauthorblockN{I Putu Arya Dharmaadi}
% \IEEEauthorblockA{\textit{Bernoulli Institute} \\
% \textit{University of Groningen}\\
% Groningen, The Netherlands \\
% arya.dharmaadi@rug.nl}
% \and
% \IEEEauthorblockN{Elias Athanasopoulos}
% \IEEEauthorblockA{\textit{Department of Computer Science} \\
% \textit{University of Cyprus}\\
% Nicosia, Cyprus \\
% athanasopoulos.elias@ucy.ac.cy}
% \and
% % \IEEEauthorblockN{Fadi Mohsen}
% % \IEEEauthorblockA{\textit{Bernoulli Institute} \\
% % \textit{University of Groningen}\\
% % Groningen, The Netherlands \\
% % f.f.m.mohsen@rug.nl}
% % \and
% \IEEEauthorblockN{Fatih Turkmen}
% \IEEEauthorblockA{\textit{Bernoulli Institute} \\
% \textit{University of Groningen}\\
% Groningen, The Netherlands \\
% f.turkmen@rug.nl}
% }

% \and

\maketitle

\begin{abstract}
Coverage feedback is an important source of guidance for fuzzing. However, obtaining such feedback %from server-side web applications 
normally requires application-level instrumentation that is specific to the language and runtime of the application.
Given that modern web applications span multiple languages and runtimes, this application-level instrumentation is costly to implement and maintain. Therefore, we present \tracelib, a system-call feedback mechanism for enabling language-agnostic web fuzzing. Our proposed approach observes the transitions of system calls, enriches selected transitions with bounded argument hashes, and converts them into a 65,536-position AFL-like bitmap. By using the generated bitmap, any web fuzzer can decide whether to retain requests that add previously unseen bitmap positions without consulting application code coverage.

We integrate TraceLib into WebFuzz and evaluate it under five WebFuzz feedback modes: the two proposed TraceLib variants (one over every traced system call and one projected onto monitored file paths and recognized SQL buffers), the N-gram comparator adapted from Xiao et al. \cite{10.5555/3766078.3766399} representing the most recent work to our knowledge, WebFuzz's Native grey-box feedback, and black-box fuzzing without feedback. 
% WebFuzz's AST edge counter serves as the measurement oracle for all five modes, but provides feedback to Native alone. 
We evaluate \tracelib on sixteen web applications under test (WUTs): eight PHP applications, and eight further applications spanning Node.js, Ruby, Java, Go and Python to demonstrate platform portability.
% In eight PHP-based Web Under Tests (WUTs) we consider, even though WebFuzz's edge counter measures every mode it guides only Native. In addition, we also evaluate six additional WUTs across Go, Python, Ruby, and Node.js to demonstrate the platform portability. 
The results show that our proposed \tracelibprojected~exceeds black-box on all eight PHP WUTs while exceeding Native on four: Joomla, Drupal, PrestaShop, and Bagisto. In addition, measured on an identical replayed request workload, the tracer costs approximately one millisecond of server-side latency per request. 
These results indicate that compact system-call feedback is a useful runtime-independent proxy for coverage guidance. Open problems include the restriction to one active request at a time, dependence on application externalization, and reduced effectiveness on encrypted database connections.

\end{abstract}

\begin{IEEEkeywords}
language-agnostic, web fuzzer, syscall, trace.
\end{IEEEkeywords}

% \begin{keywords}
%  \sep language-agnostic \sep web fuzzer \sep syscall \sep trace \sep
% \end{keywords}

% \maketitle

% \section{Introduction}
% This paper aims to determine whether system calls and SQL queries can provide meaningful feedback for guiding the prioritization and mutation of fuzzing requests. Crucially, our work focuses on evaluating the feasibility of leveraging such traces as an alternative to traditional instrumentation-based feedback mechanisms.

% This situation creates a fundamental problem: a web fuzzer only works for testing a specific web environment, and it is challenging to scale it to other environments. The main research question is: \textbf{\textit{how can we design language-agnostic feedback from the web under test while maintaining high coverage testing?}}

% Our hypothesis: we can observe system calls and submitted SQL queries for guiding the fuzzer. We believe this feedback is good for coverage guidance.

% \subsection{Main Question}
% Are system calls and SQL queries good for coverage guidance? 

\section{Introduction}
\label{intro}
Software vulnerabilities continue to pose a critical threat to modern digital infrastructure, and fuzzing has become one of the most effective and widely adopted techniques for automated vulnerability discovery \cite{yun_fuzzing_2023}. Fuzzers can trigger unexpected behaviors and crash-inducing conditions that developers may never have anticipated by automatically generating and executing large numbers of controlled random inputs \cite{zhu_fuzzing_2022}. In particular, coverage-guided fuzzing has demonstrated remarkable success in detecting memory corruption and logic bugs in native applications by leveraging lightweight feedback mechanisms, such as instrumentation-based code coverage \cite{beaman_fuzzing_2022}. Tools such as AFL \cite{zalewski2015afl}, AFL++ \cite{257204}, and libFuzzer \cite{llvmLibFuzzer} have significantly advanced the field by applying coverage feedback to guide fuzzing inputs more intelligently. This strategy has proven effective at exposing hidden bugs in complex, widely deployed software systems.

\begin{table}[t]
\centering
\caption{Comparison of grey-box web fuzzers for PHP-based web applications, our proposed \textbf{\fuzzer} with \textbf{\tracelib}, and black-box fuzzers (which are inherently language-agnostic as well).}
\label{tab:fuzz-comparison}
\begin{tabular}{lcccc}
\toprule
\multirow{2}{*}{\textbf{Fuzzer Name}} & \textbf{Internal} & \multicolumn{2}{c}{\textbf{Required Modifications}} \\ \cline{3-4} 
& \textbf{Feedback} & \textbf{Source Code} & \textbf{Interpreter} \\
\midrule

WebFuzz \cite{van_rooij_webfuzz_2021} & \ding{51} & \ding{51} & -\\ \hline
Witcher \cite{trickel_toss_2023} & \ding{51} & -& \ding{51}\\ \hline
Atropos \cite{guler_atropos_2024} & \ding{51} & - & \ding{51} \\ \hline
Phuzz \cite{neef_what_2024} & \ding{51} & - & \ding{51} \\
\bottomrule
\textbf{\fuzzer} & \ding{51} & - & -\\
\bottomrule
Black-box Fuzzers & - & - & -\\
\bottomrule

\end{tabular}
\end{table}

In the domain of web applications, grey-box web fuzzers have also emerged, which incorporate lightweight runtime feedback, typically in the form of code coverage, to guide fuzzing more effectively. These tools often instrument the source code or runtime environment of the Web application Under Test (WUT), enabling them to prioritize inputs that explore new execution paths (see Table \ref{tab:fuzz-comparison}). A notable example is WebFuzz \cite{van_rooij_webfuzz_2021}, which applies Abstract Syntax Tree (AST)-level instrumentation to PHP applications to collect code coverage and guide corpus selection. Other examples include Witcher \cite{trickel_toss_2023}, which modifies the PHP engine (i.e., interpreter) to collect code coverage to guide fuzzing, and Atropos \cite{guler_atropos_2024}, which modifies the web engine (i.e., server) in addition to the interpreter to collect all compared strings.

Although instrumentation and server-modification techniques provide effective guidance for fuzzing, they introduce significant practical limitations because they are closely tied to the programming language or runtime environment, restricting the applicability of many fuzzing frameworks to a narrow range of platforms and technologies. In applications built using multiple languages or heterogeneous technology stacks, applying instrumentation consistently across all components may require substantial engineering effort.

% \begin{table}[t]
% \centering
% \caption{Comparison of grey-box web fuzzer working for PHP-based web applications.}
% \label{tab:fuzz-comparison}
% \begin{tabular}{lcccc}
% \toprule
% \multirow{2}{*}{\textbf{Fuzzer Name}} & \multirow{2}{*}{\textbf{}} & \multicolumn{3}{c}{\textbf{Required Modifications}} \\ \cline{3-5} 
% & & \textbf{Source Code} & \textbf{Interpreter} & \textbf{Web Server} \\
% \midrule

% WebFuzz \cite{van_rooij_webfuzz_2021} & & \ding{51} & & - \\ \hline
% Witcher \cite{trickel_toss_2023} & & -& \ding{51} & - \\ \hline
% Atropos \cite{guler_atropos_2024} & & - & \ding{51} & \ding{51} \\ \hline
% Phuzz \cite{neef_what_2024} & & - & \ding{51} & - \\
% \bottomrule
% \textbf{AgnosticFuzz} &  & - & - & -\\
% \bottomrule

% \end{tabular}
% \end{table}

These limitations underscore the need for a language-agnostic feedback-driven fuzzing approach, referred to as \textbf{\fuzzer} in what follows, that preserves the advantages of grey-box fuzzing, namely guided input generation, without relying on language-specific instrumentation. Decoupling the feedback mechanism from the internal structure of the application and from its language-specific runtime enables broader applicability of the approach to diverse web applications, particularly those composed of multiple programming languages or frameworks.

To address this issue, this paper explores the possibility of leveraging externally observable runtime behaviors, specifically system-call traces, as internal execution feedback for fuzzing of web applications. By collecting system-call transitions and bounded hashes derived from selected arguments, a web fuzzer can prioritize inputs that trigger new behaviors, enabling deeper exploration of application logic in a non-intrusive and scalable manner. This design preserves the feedback-guided fuzzing loop while decoupling its guidance mechanism from the WUT’s implementation language. More importantly, it can support fuzz testing across a diverse set of web languages and execution environments.

We implement this idea in C as an eBPF-based collector named \textbf{\tracelib}.
eBPF (extended Berkeley Packet Filter) is a Linux kernel technology that allows monitoring programs to be attached to kernel hooks or events, such as those triggered when the web server performs system-call operations~\cite{ebpf}. 
Using this technology, \tracelib tracks the system calls of specific web server process IDs, converts argument-selective system-call transitions into 65,536-byte AFL-like bitmaps, and delivers the bitmaps via POSIX shared memory. Then, we attach TraceLib to WebFuzz \cite{van_rooij_webfuzz_2021} to produce \textbf{\fuzzer} by replacing WebFuzz's original source code instrumentation while preserving its overall architecture. Selecting WebFuzz as the baseline is natural because it relies solely on crawling and guided mutations to extend coverage, and its novel instrumentation captures actual edge transitions, which are suitable for evaluating syscall guidance. 

We then evaluate our proposed TraceLib against WebFuzz's Native grey-box mode, black-box mode, and another TraceLib configuration that adapts the N-gram system-call pattern coverage proposed by Xiao et al. \cite{10.5555/3766078.3766399}. To focus on feedback-guided fuzzing rather than endpoint discovery, we skip the crawling phase in WebFuzz, ensuring that every campaign receives the same 10 validated GET endpoints as seeds. Every campaign runs with 1 worker for 2 hours and is repeated three times.
% For PHP-based WUTs, WebFuzz's edge counter measures the coverage of every mode, but it guides only the Native mode. 
The evaluation includes 8 popular PHP-based WUTs and 8 non-PHP WUTs to broaden the portability study to Go, Python, Ruby, Java, and Node.js.

The results show that the best of our proposed TraceLib variants, namely \tracelibprojected, exceeds black-box on all eight PHP-based WUTs and exceeds Native on four: Joomla, Drupal, PrestaShop, and Bagisto. Averaged over the eight PHP-based WUTs and three campaigns, \tracelibprojected~reaches 11.21\% coverage, compared with 11.15\% for Native. In addition, measured on an identical replayed request workload, the tracer costs approximately one millisecond of server-side latency per request. The results support system-call feedback as a useful runtime-independent guidance proxy, while some limitations need to be addressed for future work.

\subsection*{Contributions}
In summary, we make the following contributions.
\begin{enumerate}
    \item \textbf{New Language-Agnostic Feedback Method for Web Fuzzing:}
    To provide coverage feedback for web fuzzing without relying on language-specific instrumentation, we propose a system-call-based feedback method. The method extracts transitions between selected system calls and incorporates bounded hashes of their arguments to capture web application behavior. This approach removes the need for source-code access or language-specific runtime support while remaining non-intrusive to the target application.

    \item \textbf{Analysis of Syscall-based Feedback Challenges:}
    We identify and analyze three obstacles that make raw system-call tracing difficult to use as web-fuzzing feedback: associating system calls with individual HTTP requests, bridging the abstraction gap between system-call activity and application-level behavior, and delivering feedback with sufficiently low latency (Section~\ref{sec:challenges}).

    \item \textbf{TraceLib Implementation:}
    To realize the proposed feedback method, we develop \textbf{\tracelib}, a tool that traces system calls from specific web server processes, converts the observed transitions into AFL-like coverage bitmaps, and delivers the resulting feedback through POSIX shared memory. We release the implementation of \tracelib in our public repository\footnote{\url{https://github.com/websecfuzz/tracelib}} to support reproducibility and further research.

    \item \textbf{Multi-Runtime Evaluation:}
    To assess the effectiveness and applicability of the proposed approach, we evaluate \tracelib on 16 WUTs implemented in PHP, Go, Python, Ruby, Java, and Node.js. We compare it against WebFuzz's Native grey-box mode, a black-box mode, and an N-gram system-call feedback method adapted from Xiao et al.~\cite{10.5555/3766078.3766399}. We evaluate application coverage, feedback alignment, runtime overhead, and applicability across different language runtimes.
  
  % We analyse obstacles that make the system-call tracing has not served as web fuzzing feedback, namely request association, abstraction mismatch, and delivery latency (Section~\ref{sec:challenges}), and derive mechanisms that addresses them.
  
\end{enumerate}

\section{Background}
In this section, we provide background on fuzzing and system calls. %We then summarise the internals of WebFuzz, which serves as a baseline for our work.

\subsection{Fuzzing}
Fuzzing is a software testing technique that involves automatically generating and executing large numbers of inputs to trigger unexpected behaviors, such as crashes, assertion failures, or memory errors \cite{liang_fuzzing_2018} \cite{manes_art_2021}. Basically, a fuzzer operates in an iterative loop that typically consists of three key stages: input generation, execution, and evaluation \cite{li_fuzzing_2018} \cite{dharmaadi_fuzzing_2025}. 
During input generation, the fuzzer mutates or synthesizes new test cases; during execution, these inputs are run against the target program; and in the evaluation stage, the fuzzer inspects the outcome to determine whether the input uncovered new behavior or caused a failure. Among these stages, 
the feedback gained in the evaluation step is the critical part because it steers the whole process.

\subsubsection{Feedback-guided Fuzzing}
Feedback-guided fuzzing uses runtime information (such as code coverage or application state changes) to steer input generation toward unexplored behaviors. It monitors how inputs affect execution and uses this feedback to create new, more promising inputs that reach new parts of the program. Tools like AFL insert compiler-level instrumentation into the software under test to obtain internal feedback at runtime, while others (e.g., libFuzzer and SanitizerCoverage) rely on built-in hooks from the compiler or runtime environment.

\subsubsection{Code Coverage and Bitmap File}
The coverage that is collected by the feedback-guided fuzzing is often represented as a bitmap that tracks edges or basic blocks executed during a run. A bitmap file is a data structure with a fixed-size array of bytes, often 64 KB (65,536 bytes). Each byte of the bitmap corresponds to a hashed representation of a program edge, making each entry in the bitmap function as a coarse-grained indicator of whether a particular control-flow edge or block has been observed. When a new input toggles a previously unset bitmap entry, the fuzzer interprets this as evidence of novel behavior and records the input as part of its evolving corpus.

% \subsubsection{Bitmap File}

\begin{figure}
\centering
% \begin{minted}[
%     fontsize=\small,
%     breaklines,
%     breakanywhere,
%     frame=single,
%     framesep=3mm
% ]{text}
\begin{lstlisting}
getcwd("/myproject", 4096) = 11
newfstatat(AT_FDCWD, "/myproject/.....
newfstatat(AT_FDCWD, "/usr/local/....
newfstatat(AT_FDCWD, "/myproject/wp-....
openat(AT_FDCWD, "/myproject/wp-....
newfstatat(8, "", {st_mode=S_IFREG|....
read(8, "<?php\r\n/**\r\n * Contains....
fcntl(3, F_SETLKW, {l_type=F_WRLCK, ....
fcntl(3, F_SETLK, {l_type=F_UNLCK, ....
close(8) = 0
sendto(7, "Q\0\0\0\3SELECT option_v....
poll([{fd=7, events=POLLIN|POLLERR|....
recvfrom(7, "\1\0\0\1\1D\0\0\2\3def....
\end{lstlisting}
% \end{minted}
\caption{An example of system-call traces captured from a PHP web server executing WordPress \cite{wordpress_2025}. Certain system calls provide useful information for execution path selection, such as \textit{\textbf{openat}} indicating a file being opened and \textit{\textbf{sendto}} showing submitted SQL queries.}
\label{fig:syscall-trace}
\end{figure}

% \begin{figure}
% \centering
% \begin{lstlisting}[
%     basicstyle=\small\ttfamily,
%     breaklines=true,
%     breakatwhitespace=false,
%     frame=single,
%     framesep=1mm
% ]
% getcwd("/myproject", 4096) = 11
% newfstatat(AT_FDCWD, "/myproject/.....
% newfstatat(AT_FDCWD, "/usr/local/....
% newfstatat(AT_FDCWD, "/myproject/wp-....
% openat(AT_FDCWD, "/myproject/wp-....
% newfstatat(8, "", {st_mode=S_IFREG|....
% read(8, "\<?php\r\n/\*\*\r\n \* Contains....
% fcntl(3, F_SETLKW, {l_type=F_WRLCK, ....
% fcntl(3, F_SETLK, {l_type=SIF_UNLCK, ....
% close(8) = 0
% sendto(7, "Q\0\0\0\3SELECT option_v....
% poll([{fd=7, events=POLLIN|POLLERR|....
% recvfrom(7, "\1\0\0\1\1D\0\0\2\3def....
% \end{lstlisting}
% \caption{An example of system-call traces captured from a PHP web server executing WordPress \cite{wordpress_2025}. Certain system calls provide useful information for execution path selection, such as \textit{\textbf{openat}} indicating a file being opened and \textit{\textbf{sendto}} showing submitted SQL queries.}
% \label{fig:syscall-trace}
% \end{figure}

\subsection{System Calls}
\label{syscall}
System call is a programmatic way for a computer program to request a service from the operating system's kernel, such as accessing hardware, managing files, creating or controlling processes, or communicating between programs \cite{barry_chapter_2012}. Since user-space programs are restricted from performing these operations directly, system calls act as a strictly controlled gateway between user mode and kernel mode \cite{tanenbaum2015modern}. They ensure security, isolation, and stability across processes and applications. In practice, system calls are typically invoked through wrapper functions provided by standard libraries (e.g., glibc on Linux) \cite{bagherzadeh_analyzing_2018}.

Figure \ref{fig:syscall-trace} shows an example trace of system calls executed by a PHP web server. Each request typically involves dozens to hundreds of syscalls, such as \texttt{openat}, \texttt{read}, \texttt{getcwd}, \texttt{recvfrom}, and \texttt{sendto}, reflecting the wide range of interactions between the application and the underlying operating system. When system calls are traced, such as via \texttt{ptrace} \cite{ptrace-manpage}, \texttt{strace} \cite{strace}, \texttt{eBPF} \cite{ebpf}, or custom syscall monitors, additional overhead is introduced. Each entry and exit from a syscall may generate multiple context switches between the tracer and the tracee. Even though syscall tracing can be particularly expensive, it provides an informative layer for observing program behavior.

\section{Challenges of Applying Syscall Tracing}
\label{sec:challenges}
As explained in Section \ref{intro} and Section \ref{syscall}, syscall-level tracing offers a compelling alternative to conventional code-coverage feedback for grey-box web application fuzzing, which commonly relies on code instrumentation. In addition to eliminating instrumentation-related activities, syscall tracing provides visibility into low-level behaviors that might otherwise be inaccessible through source-level instrumentation.
% % In addition to the elimination of instrumentation-related activities, syscall tracing provides visibility into low-level behaviours that might be otherwise inaccessible through source-level instrumentation.
% % the use of syscalls may enable access to information that may not be observable otherwise
% % thanks to the capture of low-level activities and signals, which might be useful feedback for fuzzing. 
% This low-level information is common across implementations and frameworks, making syscall-level tracing a promising foundation for language-agnostic feedback \cite{10.5555/3766078.3766399}.
% % promising syscall-level tracing language-agnostic coverage, which can avoid the instrumentation step. 
However, there are certain technical challenges that must be overcome to make syscall-driven feedback a practical substitute for traditional coverage metrics in grey-box web fuzzing. In what follows, we discuss at these challenges.
% However, leveraging system calls as the primary feedback mechanism introduces several technical challenges that do not arise in standard coverage-guided fuzzing.

\subsection{Challenge 1: Trace-Request Association}
\label{trace-association}
The first challenge concerns accurately attributing syscall traces to individual HTTP requests. Unlike binary fuzzing, where a single target process handles exactly one input per execution, a web server is a long-running process that services many requests over its lifetime. The fundamental difficulty is not merely collecting syscall events from the right processes, but establishing clear boundaries between the events belonging to one request and those belonging to the next. Since the tracer observes a continuous stream of syscall events from a persistent server process, there is no inherent OS-level signal that marks where the handling of one HTTP request ends and the handling of the next begins. Without such boundaries, it is impossible to reliably construct a bitmap that reflects the execution triggered by a single input.

This boundary problem is further complicated by the diversity of web server process architectures. Different server designs expose fundamentally different relationships between processes and requests. In master-worker architectures such as Apache with PHP, a master process accepts incoming connections and delegates request handling to a pool of worker processes, each of which may serve multiple sequential requests over its lifetime. In contrast, event-driven single-process architectures such as Node.js handle all requests within a single long-running process, multiplexing them through an asynchronous event loop. Between these extremes lie hybrid models such as Gunicorn and PHP-FPM, which maintain worker pools of varying sizes with different request assignment strategies. Because the mapping between processes, threads, and individual HTTP requests differs substantially across these architectures, it is challenging to design a single fixed strategy for associating syscall events with requests.

% When web servers employ parallel processes in rendering web applications, correlating and mapping of HTTP requests with their associated execution traces is challenging. Without proper server configuration and mapping of the methods, the web application fuzzers cannot obtain correct feedback from the submitted requests.

% % \subsubsection{Multi-process and Multi-threaded Applications}
% Many web frameworks adopt multi-process models, such as pre-forking workers in Python’s Gunicorn \cite{gunicorn} or process pools in PHP-FPM \cite{php}, to maximize isolation and parallelism. In addition, some frameworks rely heavily on multi-threading or event-driven concurrency, where numerous execution contexts interleave on the same worker. As a result, the syscall log is typically fragmented across multiple processes or threads, and none of them individually represents the complete behaviour of a single logical web request.

% \subsubsection{Trace Correlation and Mapping}

\subsection{Challenge 2: Information Mismatch}
\label{low-level-information}
The second challenge in adopting syscall-level tracing as feedback for web application fuzzing arises from the mismatch between the level of abstraction at which system calls operate and the logical structure of web applications. System calls capture low-level interactions such as reading from sockets, writing to files, allocating memory, or spawning threads. Although these operations are essential for program execution, they provide only indirect evidence related to the control flow or state transitions triggered by a web request. In addition, syscall sequences are shaped not only by application logic but also by runtime subsystems, such as garbage collectors, compilers, cryptography libraries, TLS stacks, and asynchronous schedulers. These components introduce additional noise that further obscures the relationship between syscall traces and the execution paths taken by the application. %Consequently, extracting meaningful, stable features from noisy, low-level events becomes a substantial challenge, often requiring specialised filtering, clustering, or modelling techniques.
% Consequently, this challenge raises a practical design question: should the fuzzer operate on the full unfiltered syscall trace, or should it apply a filter that retains only syscalls likely to reflect application-level logic?
Consequently, this challenge raises a practical design question of whether to use the full syscall trace or to retain only syscalls likely to reflect application-level logic.

Filtering the trace to semantically relevant syscalls, such as file operations (\texttt{open}, \texttt{read}, \texttt{write}, \texttt{close}), process operations (\texttt{execve}, \texttt{fork}, \texttt{clone}), network operations (\texttt{connect}, \texttt{accept}, \texttt{send}, \texttt{recv}), and inter-process communication (\texttt{socket}), may improve signal quality at the cost of potentially discarding informative events. However, whether filtered traces empirically produce better fuzzing guidance than unfiltered traces is not proven yet, since some noisy syscalls may still encode useful distinctions between execution paths. 
We therefore treat this as an empirical question and evaluate both configurations (described in Section \ref{tracelib-modes}) to determine which setting produces superior coverage guidance.

% Another related challenge is deciding how much argument information to preserve. System-call identities alone may collapse application behaviours. Retaining complete arguments, however, can introduce an impractical amount of feedback space. 

\subsection{Challenge 3: Result Delivery}
\label{result-delivery}
Once syscall-level information is collected, the feedback must be delivered back to the fuzzing engine quickly enough to guide input generation in near real time. WebFuzz \cite{van_rooij_webfuzz_2021}, the conventional web grey-box fuzzer, relies on a simple, file-based communication mechanism: the WUT writes coverage to a designated file, and the fuzzer repeatedly polls or reads that file after each test execution. 
% Although this approach is easy to implement and largely language-agnostic, because syscall traces are frequent and produced by multiple processes or threads, funnelling this information through the filesystem imposes substantial overhead due to disk I/O.
Although this approach is easy to implement and largely language-agnostic, forwarding every observed system call to userspace or writing every event to a file would introduce substantial data-transfer costs.
A practical delivery mechanism must therefore aggregate frequent events before they cross into userspace, publish one complete result for each request, and prevent the fuzzer from reading a partially updated map.

% \section{Proposed Approach}

% \subsection{Proposed Tracer}
% Given the process ID (PID) of the web server, our proposed tracer observes all syscall traces done by the PID. Following the AFL fuzzing workflow, the tracer stores syscall transitions in a bitmap file. For example, the transition of openat to read is assigned with code 1837030663, so the index of 1837030663 in the bitmap file is counted as 1. 

% \subsection{System Call Commands}
% % We observe key commands from the syscall trace to remove noise or less important information.
% We observe all commands from the syscall trace. To make it easy for further analysis, we only store the numerical forms of the commands rather than the full text. In addition, when we observe commands like openat, we also extract the file path being opened by the WUT. We think it is important to store the order of the files being executed by WUT. It is similar to code line execution, but in a higher level (i.e., file execution). 
% Furthermore, we also extract the SQL queries when the syscall records write commands.

% \subsection{SQL Query}
% % We extract operation names (e.g., SELECT, UPDATE, or DELETE) and table names from each SQL query call. 
% When receiving a SQL query, we remove the user input parts. For example, SELECT * from table WHERE name='user', we remove user, making the remaining query is SELECT * from table WHERE name. This removal aims to find unique queries generated by the WUT.

% \section{Implementation}

\section{Proposed Approach}
\label{proposed-approach}
To address the aforementioned challenges, we propose TraceLib, a language-agnostic syscall tracing library that provides per-request coverage feedback for web fuzzers without requiring any modification to the web application's source code or runtime interpreter for guidance. Ultimately, this tool guides web fuzzing by identifying requests that produce previously unexplored positions in the system-call bitmap.

% To address the aforementioned challenges, we propose TraceLib, a tool that tracks certain process IDs in a web server, converts the tracked results to be AFL-like bitmaps, and delivers the bitmaps via POSIX shared memory. Ultimately, this tool guides language-agnostic web fuzzing in evaluating submitted inputs.

% Figure \ref{fig:agnos-overview} shows the position of TraceLib in language-agnostic fuzzing.

% \subsection{TraceLib}
% Utilising the \textit{ptrace} library, TraceLib is designed to observe and monitor running processes on a Linux system (see Figure \ref{fig:agnos-overview}). Given the PID (process ID) of the web server (e.g., Apache PHP) as the target, the tool then enables a set of tracing options that allow it to observe system calls, process creation events, and program executions in the target’s process tree. The target and any of its future child processes will run under the tracer’s supervision.

% \subsubsection{Getting main PID}

% \subsubsection{Following forked process}

\subsection{TraceLib Overview}
\begin{figure}
    \centering
    \includegraphics[width=0.8\linewidth]{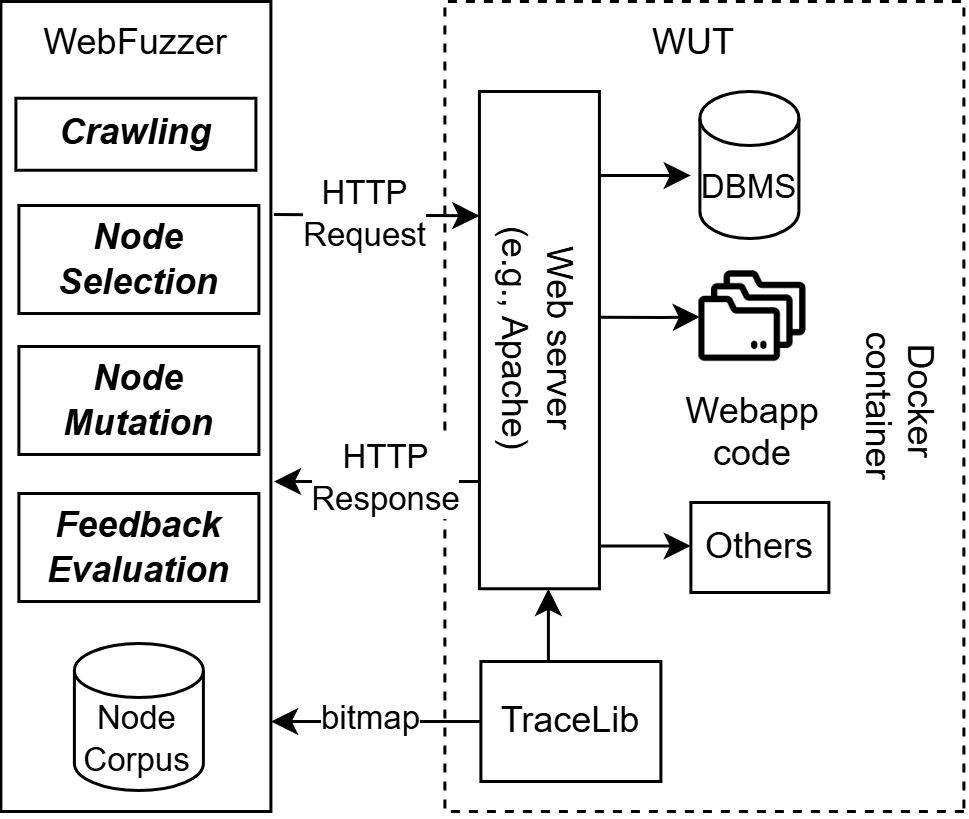}
    \caption{Web fuzzing and traceLib overview.}
    \label{fig:agnos-overview}
\end{figure}

TraceLib operates as an independent process that runs alongside the WUT during a fuzzing session. It requires only one mandatory configuration input from the user: the TCP port number on which the web server listens, together with an optional request-identifier header name (default \texttt{X-REQUEST-ID}). From this starting point, TraceLib autonomously identifies the relevant server processes, observes their system-call activity through eBPF tracepoints, and produces per-request coverage bitmaps. The fuzzer communicates with TraceLib exclusively through two channels: a custom HTTP request header that carries a unique request identifier, and the \texttt{/dev/shm} filesystem from which the fuzzer reads the resulting bitmap after each request is executed (see Figure \ref{fig:agnos-overview}).

\begin{figure}
    \centering
    \includegraphics[width=1.0\linewidth]{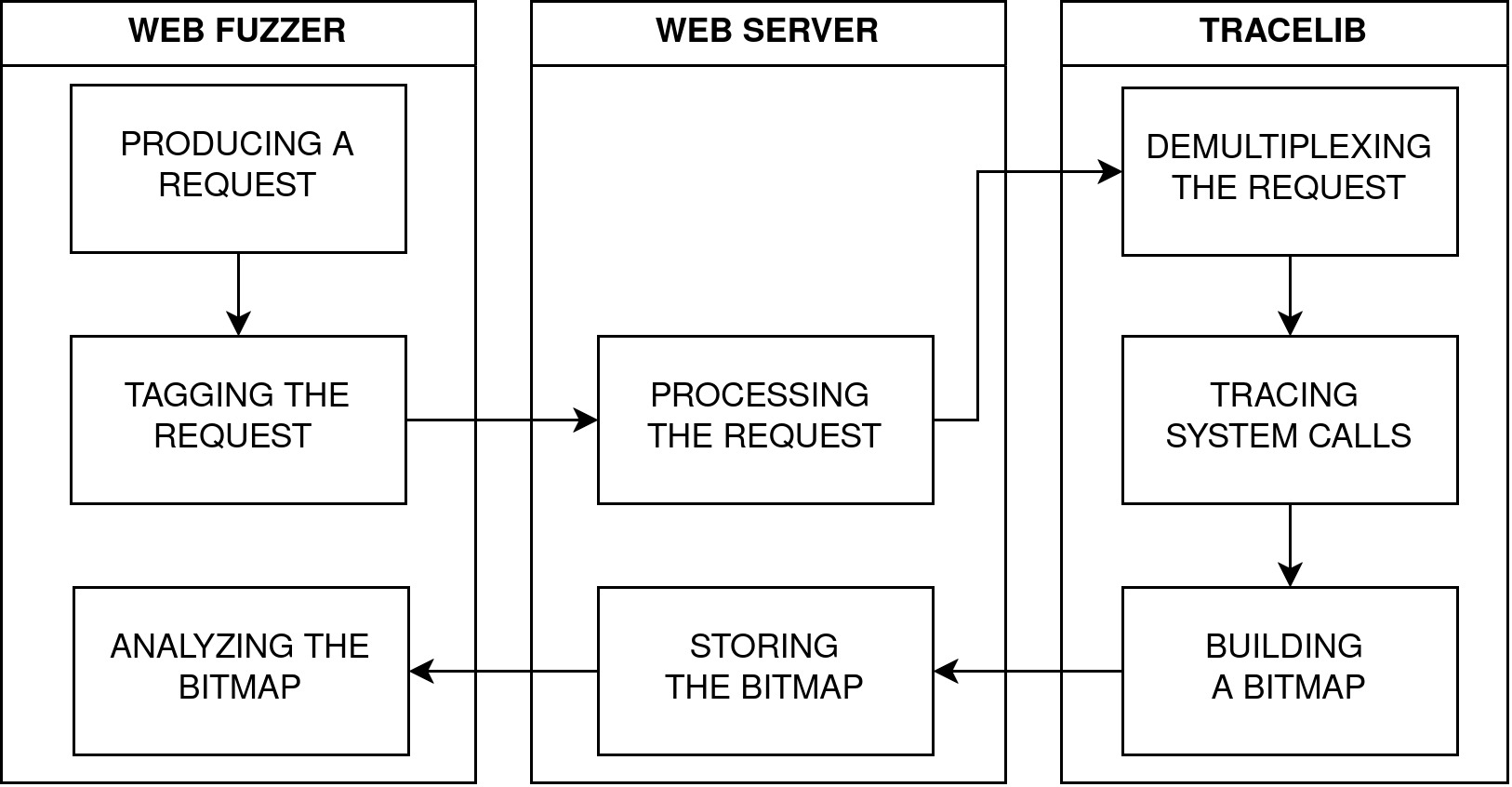}
    \caption{TraceLib request-to-feedback workflow.}
    \label{fig:tracelib-workflow}
\end{figure}

As illustrated in Figure \ref{fig:tracelib-workflow}, TraceLib has four main components: a \textit{Port-to-PID Resolver} that maps the target port to a set of server process IDs; a \textit{Tracer} that observes system calls from those process groups; a \textit{Bitmap Builder} that converts the observed syscall stream into an AFL-like coverage bitmap; and a \textit{Request Demultiplexer} that associates each bitmap with the correct fuzzer request via the HTTP header identifier. 
% While these Port-to-PID Resolver and Request Demultiplexer address the Trace-Request Association challenge (Section \ref{trace-association}) and these Tracer and Bitmap Builder address the Information Mismatch challenge (Section \ref{low-level-information}), we propose POSIX shared memory to address the result delivery challenge (Section \ref{result-delivery}). 
These components and their interactions are described in the following subsections.

% \subsection{Port-to-PID Resolver}
\subsection{Target Process Discovery}
To address the trace-request association challenge (Section \ref{trace-association}), TraceLib resolves the set of process IDs associated with the target port by cross-referencing \texttt{/proc/net/tcp} and \texttt{/proc/net/tcp6} with the file descriptor tables under \texttt{/proc/[pid]/fd/}. 
% It inserts the discovered Thread Group IDs (TGIDs) into an eBPF target map. 
% All threads of those TGIDs are then visible at the raw-system-call tracepoints, and successful \texttt{clone}, \texttt{fork}, and \texttt{vfork} returns extend the map to newly created worker processes. Sequence state is maintained per TID (Thread ID), while the positions contributed by all participating TIDs are combined in the bitmap for the active request.
The resolver identifies socket \textit{inodes} listening on the configured port and locates the processes whose file descriptors refer to those \textit{inodes}. It then inserts the corresponding thread-group identifiers (TGIDs) into an eBPF target map. While each thread has its own thread identifier (TID) in Linux, a TGID is a process-level identifier that is shared among all threads belonging to the same process.

When a web server starts, TraceLib attaches eBPF programs to \texttt{raw\_syscalls:sys\_enter} and \texttt{raw\_syscalls:sys\_exit}, two kernel tracepoints that fire whenever any process enters or exits the kernel to perform a system call. Each tracepoint invocation checks whether the current TGID is present in the target map before performing request detection or feedback collection. Because the filter is applied by TGID, system calls from all threads belonging to a tracked process are observed without attaching to every TID individually.

When a tracked process successfully executes \texttt{clone}, \texttt{fork}, or \texttt{vfork}, TraceLib inserts the returned child identifier into the target map. This mechanism keeps newly created server workers visible for tracking. Consequently, TraceLib can monitor system-call activities from both the original server process and its subsequently created workers.

% TraceLib then attaches to each discovered process via the \texttt{ptrace} interface, with options that cause tracing to be inherited during \texttt{fork}, \texttt{vfork}, and \texttt{clone} events. This mechanism directly addresses the dynamic process lifecycle concern raised in Section~\ref{low-level-information}: when a server such as PHP-FPM or Apache prefork spawns a new worker to handle a request, the kernel automatically delivers a trace event for the new child, ensuring that no part of request-handling execution escapes observation. In addition, to allow tracing threaded servers, such as Puma (Ruby) and the Go runtime's thread pool, TraceLib walks \texttt{/proc/[pid]/task/} for each discovered process and attaches every thread (TID) individually.

\begin{figure}
\centering
% \begin{minted}[
%     fontsize=\footnotesize,
%     breaklines,
%     breakanywhere,
%     frame=single,
%     framesep=3mm
% ]{text}
\begin{lstlisting}
getcwd ---> newfstatat  = 20230 [3]
newfstatat ---> newfstatat = 1542 [120]
newfstatat ---> open(home.php) = 27411 [2]
open(home.php) ---> open(main.php) = 16879 [2]
open(main.php) ---> sendto(SELECT * FROM data) = 28108 [1]
\end{lstlisting}
% \end{minted}
\caption{Illustration of how system-call transitions are recorded. Using the concatenation method in Section \ref{tracing-monitoring}, TraceLib assigns the ID of 20230 to the transition from \textit{\textbf{getcwd}} to \textit{\textbf{newfstatat}}. Therefore, in the bitmap file, the index of 20230 contains the value 3, which is the number of times this transition occurs. When a syscall brings a useful argument, such as \textit{\textbf{open}}, the argument is included in the hash calculation (explained further in Section \ref{argument-selective-syscall-transition}).}
\label{fig:tracing-illustration}
\end{figure}

\subsection{Tracing and Behavioral Monitoring}
\label{tracing-monitoring}
To address the information mismatch challenge (Section~\ref{low-level-information}), TraceLib observes the system-call transitions executed by the target processes identified in the previous section. Each resulting transition is then assigned a 16-bit \textbf{\textit{transition ID}} (TrID) and accumulated in an AFL-style coverage map consisting of 65{,}536 saturating eight-bit counters.

TraceLib produces the transition ID by concatenating the previous and current system-call numbers. Each call number, which commonly uses 9 bits in the x86-64 architecture, must be reduced to 8 bits in order to fit a 16-bit transition ID. 
Then, each 8-bit component additionally carries an 8-bit hash of that call’s argument, for the reason given in Section~\ref{argument-selective-syscall-transition}.
Therefore, given $n_i$ and $n_j$ as the syscall numbers of syscalls $i$ and $j$, respectively, the TrID from syscall $i$ to syscall $j$ is defined as:
\begin{equation}
\operatorname{TrID}(i,j) =
\hat{n}_i \,||\, \hat{n}_j,
\qquad
\hat{n}_i = (n_i \bmod 256) \oplus \alpha_i
\label{eq:trid}
\end{equation}

where $||$ denotes concatenation, $\oplus$ denotes bitwise XOR, and $\alpha_i$ is the 8-bit argument hash of call $i$, defined in Section~\ref{argument-selective-syscall-transition}.

% idx = ((prev_sc % 256) ^ prev_arg_hash)   << 8) ^ ((curr_sc % 256) ^ curr_arg_hash)

This concatenation method is different from AFL-based fuzzers, which use the XOR operation for generating an ID for block transitions. AFL can afford XOR because it assigns every basic block a random identifier, so that the randomness spreads pairs evenly over the map. On the other hand, syscall numbers are dense, small integers from a narrow range, for which an XOR produces collisions immediately. For example, a transition from call~1 (bit: 01) to call~2 (bit: 10) and one from call~3 (bit: 11) to call~0 (bit: 00) would address the same cell, namely bit 11.

\subsubsection{Argument-Selective System-Call Transitions}
\label{argument-selective-syscall-transition}
To produce an accurate coverage map, the use of transitions between system calls alone is too coarse for web workloads. Based on our observations, a request commonly performs thousands of \texttt{openat} and \texttt{write} calls whose meaning often depends on the arguments used in the call. TraceLib therefore includes argument information in the TrID calculation (Equation \ref{eq:trid}), but restricts it to the two argument families that are both semantically meaningful and stable across repetitions of the same request: file paths and SQL statements.

% More formally, after being reduced to 8 bits, each syscall number $s$ is XOR-ed with $\alpha_i$ 
% \begin{equation}
%   \label{eq:alpha}
%     \alpha_i =
%     \begin{cases}
%       h\bigl(p_i\bigr), & \text{$s_i$ opens or executes a file},\\
%       h\bigl(r(w_i)\bigr), & \text{$s_i$ writes a SQL query},\\
%       0, & \text{otherwise}, 
%     \end{cases}
%   \end{equation}

% where $p_i$ is the path argument of \texttt{open}, \texttt{openat}, \texttt{openat2}, or \texttt{execve}; $w_i$ is the buffer of \texttt{write}, \texttt{writev}, or \texttt{sendto}; $r(\cdot)$ is the query reduction (described below in Section \ref{reducing-sql-query}); and $h(\cdot)$ is a one-byte hash function.

More formally, the following hashes are used to calculate $\alpha$ in  Equation (\ref{eq:trid}) is:
\begin{equation}
\alpha_i =
\begin{cases}
h(\kappa(p_i)), & s_i \text{ opens or executes a file}, \\
h(c(w_i)), & s_i \text{ writes a SQL query}, \\
0, & \text{otherwise},
\end{cases}
\label{eq:alpha}
\end{equation}
where $p_i$ is the path argument of \texttt{open}, \texttt{openat}, \texttt{openat2}, or \texttt{execve}; $w_i$ is the buffer of \texttt{write}, \texttt{writev}, or \texttt{sendto}; $\kappa(\cdot)$ is the path policy presented in  Section~\ref{sec:path-policy}; $c(\cdot)$ is the compact query encoding presented in Section~\ref{reducing-sql-query}; and $h(\cdot)$ is a one-byte hash function using \textit{djb2} method \cite{ROUSSEV2007105}.

% The argument hash enters the transition ID by XOR, folded into the same byte as the syscall number it belongs to. It is convenient to name the result: we call
% \begin{equation}
% \hat{n}_i = (n_i \bmod 256) \oplus \alpha_i
% \label{eq:nhat}
% \end{equation}
% the \emph{argument-keyed identity} of call $i$. Both terms are 8 bits, so $\hat{n}_i$ is 8 bits, and it occupies exactly the position that $(n_i \bmod 256)$ occupied in Equation~\eqref{eq:trid-base}. A call with no argument has $\alpha_i = 0$ and therefore $\hat{n}_i = n_i \bmod 256$: the keyed identity degenerates to the bare syscall number, which is what makes Equation~\eqref{eq:nhat} a refinement of Equation~\eqref{eq:trid-base} rather than a replacement for it.

\subsubsection{File Path Filtering and Canonicalization}
\label{sec:path-policy}
Our observations of a system-call activity show that web applications often access files that are unrelated to the application code. For example, a web request may invoke \texttt{openat("/tmp/cache/sess\_a91f2c.tmp")}, where the cache file name may be generated differently for each execution. Consequently, two identical requests that run the same application code and follow the same execution path can produce different system-call traces solely because these dynamically generated file names differ. Therefore, TraceLib provides three mechanisms for controlling file-path monitoring. Together they define the function $\kappa(\cdot)$ of Equation (\ref{eq:alpha}), which returns the path that reaches the hash or nothing at all. 

The \texttt{TRACELIB\_FILE\_PATH\_MONITORED} environment variable specifies the directory whose file accesses should be monitored, with \texttt{/var/www} as the default value. When this variable is set, only file accesses to paths under the specified directory are included in transition calculations. The default suits PHP deployments; for the other runtimes, the monitored root is set per application to the directory the running process actually reads from. For example, a Ruby-based application is set to monitor files at \texttt{/usr/src}.

In addition, \texttt{TRACELIB\_EXCLUDED\_FILE\_PATH} specifies a file path or path component to exclude from monitoring, with \texttt{temp} as the default value. Thus, accesses matching the excluded path are ignored even when they fall within the monitored directory. 

The last mechanism is canonicalization, applied to every surviving path before it is hashed. Its purpose is to suppress noise caused by path components that are generated dynamically (and even randomly) and may therefore differ across executions of the same request, even when the underlying application behavior is identical. To remove this noise, a path component that is entirely numeric is replaced by \#. In addition to the numeric path, a path component that has eight or more hexadecimal characters (e.g., content-addressed names, cache keys), or sixteen or more alphanumerics that mix case and contain a digit (e.g., session identifiers, framework tokens) is replaced by \#.

% \subsubsection{SQL Query Reduction}
% \label{reducing-sql-query}
% Hashing a SQL query verbatim would be ineffective since two executions of the same page may issue the same query with different values resulting with a unique hash, and therefore every value would look like a new behaviour. The reduction function $c(\cdot)$ in Equation (\ref{eq:alpha}) removes exactly this noise. It scans the buffer for the first genuine SQL command  (e.g., \texttt{SELECT} or \texttt{INSERT}), keeps the query text from that command onwards, and deletes only the \emph{values} that appear on the right-hand side of a comparison operator (e.g., \texttt{=} or \texttt{LIKE}). 

% For example,
%   \begin{quote}\ttfamily\small
%   SELECT * FROM users WHERE id = 7 AND state LIKE 'ok'\\
%   \end{quote}

% is reduced to be:
%   \begin{quote}\ttfamily\small
%   SELECT * FROM users WHERE id = AND state LIKE
%   \end{quote}
  
% By using this method, the same query with a different user id yields the same reduced query, whereas a query that touches another table, selects other columns, or uses another operator produces a different one, which is precisely the distinction a fuzzer should react to. 

\subsubsection{Compact SQL Query Encoding}
\label{reducing-sql-query}
Hashing a SQL query verbatim would be ineffective since ten executions of the same code may issue queries with different values resulting with a unique hash, and therefore every value would look like a new behavior. The compact function $c(\cdot)$ in Equation (\ref{eq:alpha}) is designed to remove this noise. Instead of preserving the entire query, it extracts only two components that are important for identifying the database operation: the SQL commands and the table names.

For example, a query of
  \begin{quote}\ttfamily\small
  SELECT p.* FROM posts p LEFT JOIN postmeta m ON m.post\_id = p.ID WHERE p.author\_id = 7\\
  \end{quote}

is reduced to be:
  \begin{quote}\ttfamily\small
  <<SELECT,LEFT JOIN>><<POSTS,POSTMETA>>
  \end{quote}

We design this compact query approach to be simple to implement. Rather than parsing the full SQL grammar, it recognizes SQL commands using a predefined dictionary and extracts table names based on the keywords that introduce them, such as \texttt{FROM}, \texttt{JOIN}, \texttt{UPDATE}, and \texttt{INTO}. All other parts of the query are discarded. This design keeps the encoder lightweight while retaining information about which database operations and tables are involved.

\paragraph*{Alternative Query Reduction}
We also considered a more direct reduction strategy that removes only the user values from the query while preserving the rest of the statement. Although this strategy can remove value-dependent noise, it is harder to implement. First, user values can appear in many different positions within a query, making it difficult to define general rules that identify and remove them. Second, complex web applications often construct queries from nested variables, so the textual structure may change even when the underlying database operation is logically the same.

For example, an array used in an \texttt{IN} clause may produce \texttt{IN (7)} for one element and \texttt{IN (7, 8, 9)} for three elements. Even after removing the values, the number of commas differs, causing the same web code to produce different hashes. We therefore use the compact representation above, which focuses on SQL commands and table names rather than attempting to normalize the full query text.

% \subsubsection{N-Gram Comparator}
% \subsubsection{Two TraceLib Modes and N-Gram Comparator}
\subsubsection{Two TraceLib Modes}
\label{tracelib-modes}
For a more comprehensive evaluation, we create two TraceLib modes that differ in which system calls are included in the transition calculation: \textbf{\textit{\tracelibplain}} and \textbf{\textit{\tracelibprojected}}. Both modes use the same encoding approach described above. However, \textbf{\textit{\tracelibplain}} includes all system calls in the transition calculation, regardless of their arguments, whereas \textbf{\textit{\tracelibprojected}} records only system calls for which $\alpha_i \neq 0$. The motivation for \textbf{\textit{\tracelibprojected}} is to approximate application-level code behavior by focusing on transitions involving semantically meaningful resources, such as opening files and executing SQL queries.

\subsubsection{N-Gram Comparator}
\label{n-gram-comparator}
In addition, to provide a meaningful comparison with prior work on system-call feedback, we also create an N-gram comparator adapted (named \textbf{\textit{\tracelibngram}}) from the system-call pattern coverage proposed by Xiao et al. \cite{10.5555/3766078.3766399}. Although their approach was proposed for binary applications rather than web applications, the underlying feedback representation is directly relevant to our work.
To ensure a consistent comparison, we adopt its N-gram representation and implement it within the same tracing framework as TraceLib, rather than using the exact code of the original implementation. Specifically, this N-gram comparator uses the same process discovery, eBPF tracepoints, request demultiplexer, argument hash function, map size, and bitmap path as TraceLib. Therefore, both approaches use the same mechanisms to identify and trace target processes; they differ only in how they represent observed system-call behavior as feedback.

The N-gram comparator represents each system-call event as a 16-bit token composed of the 8-bit system-call number and the 8-bit argument hash. Unlike the argument-selective approach used by TraceLib, it incorporates all available argument data when computing the argument hash. Moreover, rather than using transitions between two consecutive system calls as in TraceLib, the N-gram comparator hashes a window of the four most recent tokens and a window of the eight most recent tokens to produce TrID.

% \subsubsection{Global Bitmap Comparison}
% Following the idea of AFL-style fuzzers, a web fuzzer utilizing TraceLib compares each completed request bitmap with the global bitmap. The global bitmap is updated when a completed request bitmap brings a new transition ID or a higher transition frequency. Then, the corresponding request is stored in the Corpus.

\subsection{Request Demultiplexing}
\label{request-demultiplexing}
To address the trace-request association challenge (Section \ref{trace-association}), TraceLib associates each bitmap with its originating HTTP request through a cooperative mechanism between web fuzzer and TraceLib. The fuzzer embeds a unique identifier in a designated HTTP header (default \texttt{X-REQUEST-ID}) of every request. Because the identifier is carried within the HTTP request itself, we refer to this as an \emph{in-band request identifier}.
% TraceLib observes this header when the web server writes or forwards the request internally, uses the value as the name of a fresh bitmap, and records all subsequent syscall events into it until the response is emitted or the next HTTP header is recognized. 
TraceLib then observes this header on both the inbound path, by scanning \texttt{read}/\texttt{recvfrom} payloads at syscall exit, and the outbound path, by scanning \texttt{write}/\texttt{send}/\texttt{sendto}/\texttt{writev} payloads at syscall entry. While the inbound scan is crucial for typical servers that process the request but do not include that custom header in their HTTP response sent back to the client, the outbound scan covers frameworks that echo the custom header back. 
TraceLib uses the custom header value as the name of a fresh bitmap and records all subsequent syscall events into it until the response is emitted or the next HTTP header is recognized.
The finalized bitmap is written to \texttt{/dev/shm} under the request identifier, where the fuzzer retrieves it by name.

\subsubsection{Non-concurrent Request Requirement}
\label{non-concurrent-request-req}
% To guarantee that an active bitmap is not contaminated by other concurrent requests, TraceLib requires the web fuzzer to send one request at a time (a.k.a a sequential request). 
To ensure that the active bitmap contains feedback attributable to a single request, TraceLib requires the web fuzzer to issue requests sequentially, with at most one request being processed at a time. Concurrent requests could cause system-call transitions from different requests to be recorded in the same bitmap, making it impossible to reliably associate the collected feedback with an individual request.
This design is consistent with the isolated-execution model of common binary fuzzers such as AFL.

% \subsubsection*{Challenges with the prefork-child-as-request Approach}
\subsubsection{Alternative Prefork-Child Association}
Another possible way to address the trace-request association challenge is to configure Apache \texttt{mpm\_prefork} with \texttt{MaxConnectionsPerChild 1}, which makes each worker process serve exactly one request and then die. The request to which a bitmap belongs is then recovered from the Apache access log, which maps the \texttt{request\_ID} in the header to the \texttt{process\_ID}. 

This technique allows the web fuzzer to send concurrent requests; however, it is structurally tied to the Apache schema, which implements a master-worker architecture. In contrast, event-driven single-process architectures, such as Node.js, handle all requests within a single long-running process and multiplex them through an asynchronous event loop, making the prefork-child-as-request approach unsuitable.
% Therefore, our design to use the in-process header demux approach is better because it is more server-agnostic.

Therefore, rather than adopting the prefork-child mechanism, TraceLib uses the \emph{in-band request identifier} with the non-concurrent request requirement as its primary design because this approach does not depend on a particular web-server process model and is more broadly applicable across different server architectures.

\subsubsection{Start-Stop Syscall Recording}
Since the kernel provides no information about HTTP request boundaries, TraceLib needs to identify the beginning and end of each request from the data exchanged by the server. As explained before, TraceLib detects the request ID in incoming data to mark the start of a request. When a new ID is detected, any previously active request is finalized, and a new coverage map is opened. If the same ID is seen again, the existing request remains active.

To detect the end of a request, we define the following rules. First, TraceLib can detect the beginning of an HTTP response. By default, the current configuration treats the first \texttt{HTTP/1.} status line as the end of the active request. Second, TraceLib uses a 500 ms idle timeout. If the server becomes completely inactive for this period, the current request is finalized. The timer is based on activity across the traced server rather than on a single thread. Finally, a request ends when a new request ID is detected or when the traced process shuts down.

\subsection{In-Kernel Aggregation and Bitmap Delivery}
To address the result delivery challenge described in Section~\ref{result-delivery}, TraceLib performs feedback aggregation in the kernel and delivers completed bitmaps to userspace through a double-buffered eBPF map and the tmpfs-backed \texttt{/dev/shm} filesystem. The map consists of two 65,536-byte halves, allowing one half to collect feedback for the active request while userspace processes the completed bitmap in the other half. Each observed system call updates the active half of the BPF map in kernel context, avoiding the need to transfer individual system-call events to userspace. When a request is finalized, TraceLib switches to the other half of the map and copies the completed 65,536-byte bitmap to a file named \texttt{/dev/shm/<request-id>}.

% Answering the last challenge described in Section \ref{result-delivery}, which is about result delivery, TraceLib uses a double-buffered eBPF map together with the tmpfs-backed \texttt{/dev/shm} filesystem. The map contains two 65,536-byte halves, allowing one half to collect the active request while userspace finalises the other. Each observed system call increments the active half of the BPF map in kernel context. When a request is finalised, the userspace collector copies the completed 65,536-byte half to a file named \texttt{/dev/shm/<request-id>}.

% \subsection{Deployment Considerations}
% We choose \emph{ptrace} over alternative tracing mechanisms, notably \emph{eBPF}, for research deployment consideration. eBPF-based tracing requires either root privileges or the \texttt{CAP\_BPF} capability, neither of which is available in standard unprivileged Docker containers. \texttt{ptrace}, by contrast, requires only the \texttt{SYS\_PTRACE} capability, which Docker enables by default. This makes TraceLib deployable in the containerised experimental environments that are increasingly expected by artifact evaluation committees, without requiring any modification to the container security profile.

\section{Agnostic Web Fuzzer Framework}
To make TraceLib usable in practice, we design an agnostic web fuzzing framework by combining TraceLib's language-agnostic coverage feedback with an existing coverage-guided web fuzzer. We first motivate our choice of base fuzzer, then describe its original architecture and the targeted modifications required to replace its native instrumentation with TraceLib.

\subsection{Rationale for a Whole-Application Base}
\label{webfuzz-rationale}
Existing grey-box web fuzzers fall into two broad categories that address fundamentally different research questions. First, \emph{per-endpoint fuzzers}, such as Witcher~\cite{trickel_toss_2023}, Atropos~\cite{guler_atropos_2024}, and Phuzz~\cite{neef_what_2024}, take a single URL or script as input and exercise it intensively for a fixed time budget. Each endpoint is fuzzed with its own configuration and fuzzer instance, to explore one code path as thoroughly as possible. The fuzzer cannot freely explore all routes and endpoints in a single, shared, app‑wide session. On the other hand, \emph{whole-application fuzzers}, such as WebFuzz~\cite{van_rooij_webfuzz_2021}, treat the web application as a complete system: they crawl its endpoints dynamically, maintain a corpus of requests spanning multiple handlers, and allocate fuzzing effort across the entire application based on accumulated coverage feedback. Their goal is to measure and improve how broadly a fuzzer can navigate a realistic application surface.

Since these two categories are different, a direct head-to-head comparison would conflate two unrelated capabilities. For the research question addressed in this paper, which is whether a language-agnostic coverage signal can effectively guide fuzzing across heterogeneous web applications, whole-application coverage is the more informative evaluation metric, since it reflects the fuzzer's ability to explore applications of realistic complexity rather than isolated scripts selected in advance. We therefore adopt a whole-application fuzzer as the base for our Agnostic Fuzzer. Within this category, WebFuzz is the natural choice.

\subsection{WebFuzz Architecture}
WebFuzz \cite{van_rooij_webfuzz_2021} is a grey-box web fuzzer, which maintains a corpus of HTTP requests and iteratively mutates them to explore the application. At startup, WebFuzz crawls the WUT to discover its endpoints and seed the corpus with requests that reach each one. During fuzzing, it repeatedly selects a request from the corpus, applies mutation operators to its parameters and payloads, issues the mutated request, and collects coverage feedback from the server. Requests that trigger previously unseen coverage are added to the corpus, and the selection probability for each corpus entry is weighted by its historical contribution to coverage growth.

In its original form, WebFuzz obtains coverage through AST-level instrumentation of the WUT. This instrumentation ties WebFuzz to PHP and requires the application's source code to be available and rewritable.

\subsection{Integrating TraceLib with WebFuzz}
Agnostic web fuzzer is achieved by replacing WebFuzz's AST-based coverage collection with TraceLib, while leaving the remainder of the fuzzer unchanged. This means that after issuing a request, WebFuzz retrieves the corresponding bitmap from \texttt{/dev/shm} under the request's identifier and compares it with the global bitmap to determine whether the request brings new behaviors or not. When new behavior is detected, the subsequent steps remain identical to the original WebFuzz: updating the corpus and performing the next fuzzing iteration.

\begin{table*}[t]
\centering
\small
\caption{Web Under Tests (WUTs) with their runtime environments, GitHub repository names, coverage counter tools used, GitHub popularity, and approximate codebase size.}
\label{tab:applications}
\setlength{\tabcolsep}{5pt}
\renewcommand{\arraystretch}{1.08}
\begin{tabular*}{\textwidth}{
    @{\extracolsep{\fill}}
    r l l l l r r
}
\toprule
\# & WUT & Runtime &
GitHub Repo Name &
\makecell{Coverage\\Counter} &
\makecell{GitHub\\Stars} &
LoC \\
\midrule
1  & WordPress   & PHP     & wordpress/wordpress             & WebFuzz Edge Counter & $\sim$21.3k & $\sim$1M     \\
2  & HotCRP      & PHP     & kohler/hotcrp                   & WebFuzz Edge Counter & $\sim$435   & $\sim$254.3k \\
3  & phpBB       & PHP     & phpbb/phpbb                     & WebFuzz Edge Counter & $\sim$2.1k  & $\sim$327.0k \\
4  & Joomla      & PHP     & joomla/joomla-cms               & WebFuzz Edge Counter & $\sim$5.1k  & $\sim$334.2k \\
5  & Bagisto     & PHP     & bagisto/bagisto                 & WebFuzz Edge Counter & $\sim$28.1k & $\sim$512.0k \\
6  & Drupal      & PHP     & drupal/drupal                   & WebFuzz Edge Counter & $\sim$4.3k  & $\sim$1.06M  \\
7  & PrestaShop  & PHP     & PrestaShop/PrestaShop           & WebFuzz Edge Counter & $\sim$9.2k  & $\sim$2.28M  \\
8  & ZenCart     & PHP     & zencart/zencart                 & WebFuzz Edge Counter & $\sim$425   & $\sim$202.4k \\
\midrule
9  & Ghost       & Node.js & TryGhost/Ghost                  & c8 (V8 coverage)       & $\sim$55.3k & $\sim$1.03M  \\
10 & Wiki.js     & Node.js & requarks/wiki                   & c8 (V8 coverage)       & $\sim$28.9k & $\sim$58.0k  \\
11 & Redmine     & Ruby    & redmine/redmine                 & Ruby \texttt{Coverage}            & $\sim$6.0k  & $\sim$219.0k \\
12 & Huginn      & Ruby    & huginn/huginn                   & Ruby \texttt{Coverage}            & $\sim$49.9k & $\sim$51.4k  \\
13 & Roller      & Java    & apache/roller                   & JaCoCo               & $\sim$133   & $\sim$77.0k  \\
14 & PetClinic   & Java    & spring-projects/spring-petclinic & JaCoCo              & $\sim$9.5k  & $\sim$12.8k  \\
15 & Superset    & Python  & apache/superset                 & Coverage.py          & $\sim$74.7k & $\sim$1.34M  \\
16 & Gogs        & Go      & gogs/gogs                       & Go Cover             & $\sim$47.8k & $\sim$140.2k \\
\bottomrule
\end{tabular*}
\end{table*}

\section{Evaluation}
In this section, we evaluate TraceLib to determine whether its system-call feedback can practically replace code instrumentation for guiding web fuzzing. 
Our evaluation is driven by the following research questions:
\begin{itemize}
    % \item How accurately does TraceLib distinguish between requests that use different application code coverages, while treating requests with identical code coverages as equivalent?
    % \item RQ1. How good is TraceLib compared to standard black-box and grey-box web fuzzing, in terms of code coverage?
    \item RQ1. How much application code does WebFuzz cover when corpus guidance comes from TraceLib rather than WebFuzz's Native instrumentation or a black-box approach?
    \item RQ2. How accurately does the feedback provided by \tracelib agree with WebFuzz's Native instrumentation and other code coverage tools?
    \item RQ3. How much time overhead does TraceLib introduce?
    % \item RQ4. How much does each of TraceLib's components contribute to the signal that drives coverage-guided fuzzing?
    \item RQ4. How portable is TraceLib across different web platforms?
    
    % \item RQ2. How many maximum concurrent requests can be handled by the server that can still give high code coverage?
    % \item RQ4. How good is its scalability and practicality, in the sense of how many web server environments can be handled with a little change of configuration/testing code?
\end{itemize}

\subsection{Evaluation Setup}
We prepare five feedback sources for this evaluation:
\begin{itemize}
    \item \textbf{Native or Grey-box}: WebFuzz's original AST-level instrumentation. This is the only configuration that requires access to the application's source code. It is available for PHP-based WUTs, but not for non-PHP WUTs because WebFuzz does not support instrumentation of their source code.
    \item \textbf{\tracelibplain} and \textbf{\tracelibprojected}: our proposed approaches, described in Section~\ref{tracelib-modes}.
    \item \textbf{\tracelibngram}: the N-gram encoding comparator adapted from \cite{10.5555/3766078.3766399}, described in Section~\ref{n-gram-comparator}.
    \item \textbf{Black-box}: a configuration that uses no feedback and retains every request for Corpus.
\end{itemize}

% Described in Section~\ref{tracelib-modes}, we evaluate three TraceLib configurations: \textbf{\textit{\tracelibplain}}, which uses all observed system calls; \textbf{\textit{\tracelibprojected}}, which retains only system calls carrying semantically meaningful arguments; and \textbf{\textit{\tracelibngram}}, which uses N-gram system-call patterns as a comparator. We compare these configurations primarily against WebFuzz's Native instrumentation and the black-box approach without feedback.

% To demonstrate proper evaluation to answer all RQs, we set up several things as follows.

% \subsubsection{TraceLib Versions}
% \label{two-tracelib-version}
% This evaluation distinguishes two TraceLib feedback variants. The first is our proposed TraceLib, described in Section \ref{proposed-approach}. The second version, which we call TraceLib with N-gram, adapts the N-gram System Call Pattern Coverage approach of Xiao et al. \cite{10.5555/3766078.3766399}. We include this variant because both approaches derive feedback from system-call behavior but encode execution context differently. TraceLib with N-gram hashes syscall contexts of length $\theta=4$ and $2\theta$, as well as the whole-request trace, into the same bitmap. The additional sequence context may distinguish meaningful execution phases, but it may also capture incidental differences from scheduling, caching, or other activities. In a nutshell, this variant is a comparative adaptation rather than an exact reproduction of the original implementation.

% \subsection{WUT and Experiment Configuration}
\subsubsection{WUT Selection}
\label{wut-selection}
To evaluate \tracelib in realistic settings, we selected 16 popular web applications (see Table \ref{tab:applications}) spanning six runtime environments (i.e., PHP, Go, Python, Ruby, Java, and Node.js). PHP-based WUTs dominate because the base fuzzer (i.e., WebFuzz) was originally designed for PHP and includes AST-level instrumentation, enabling a more detailed code coverage comparison.

\subsubsection{Code Coverage Oracle} 
\label{sec:code-coverage-oracle} 
As shown in Table \ref{tab:applications}, for PHP-based WUTs, we use WebFuzz's Edge counter as the coverage oracle because it records transitions between code blocks, rather than only the set of executed lines. Thus, the execution of a request is represented as a set of $(\textit{predecessor},\textit{successor})$ pairs, providing a more fine-grained representation of web application behavior. Because WebFuzz's Edge counter is available only for PHP, PHP-based WUTs constitute the larger part of our evaluation and allow us to perform more comprehensive coverage comparisons. 

For non-PHP WUTs, which are not supported by WebFuzz's Edge counter, we use the coverage tools provided for their respective runtimes: \texttt{go cover} for Go, \texttt{coverage.py} for Python, \texttt{JaCoCo} for Java, Ruby's built-in coverage support for Ruby, and \texttt{c8} for Node.js. These tools primarily represent coverage as a set of executed lines without preserving their execution order. Therefore, their coverage measurements have a different granularity from the edge-based measurements used for PHP and should not be interpreted as directly equivalent.

\subsubsection{Fuzzer Setup}
All campaigns are driven by the same fuzzer (webFuzz), which runs in five feedback modes. %(TraceLib, Native, Black-box). 
Single-worker operation (achieved by putting \texttt{-w 1}) is used to ensure that only one HTTP request is submitted at a time, which complies with the requirement of TraceLib's demultiplexer (see Section \ref{request-demultiplexing}). %Since eight of the WUTs are PHP-based, they are eligible for WebFuzz's Native mode; the remaining five are ineligible because the WebFuzz instrumentor is PHP-only.
% % As explained in Section \ref{webfuzz-rationale}, the crawling phase starts first, and then WebFuzz continues to the fuzzing phase after the crawling is done. Rather than using a time-based limit, we set a fuzz-request budget, which is 10K mutated requests to submit by WebFuzz, because each WUT has a varying processing time. 
% % As explained in Section
% We skip WebFuzz's crawling phase to remove endpoint discovery as a confound and to give the same start for each mode. Therefore, every mode receives the same ten authenticated GET seeds (see Appendix), each validated with an HTTP 200 response, and has a 7,200-second wall-clock budget for the fuzzing campaign. 

\subsubsection{Experiment Environment}
All experiments were conducted on an Intel Core i7-6700 CPU @ 3.40 GHz (four cores and eight threads), with 32 GiB of RAM and two 238.5 GB SSDs. The operating system is Ubuntu 26.04 LTS (64-bit), running Docker Engine 29.5.3 and Docker Compose 5.1.4 for containerized deployments. Each WUT is deployed with its runtime (Table \ref{tab:applications}). WebFuzz runs on the host in a Python 3.12.13 virtual environment.

% \begin{figure*}[t]
%     \centering

%     % Row 1
%     \begin{subfigure}{0.32\textwidth}
%         \centering
%         \includegraphics[width=\linewidth]{data/wordpress.png}
%         \caption{WordPress}
%     \end{subfigure}
%     \hfill
%     \begin{subfigure}{0.32\textwidth}
%         \centering
%         \includegraphics[width=\linewidth]{data/drupal.png}
%         \caption{Drupal}
%     \end{subfigure}
%     \hfill
%     \begin{subfigure}{0.32\textwidth}
%         \centering
%         \includegraphics[width=\linewidth]{data/zencart.png}
%         \caption{Zencart}
%     \end{subfigure}

%     \vspace{0.5em} % Space between rows

%     % Row 2
%     \begin{subfigure}{0.32\textwidth}
%         \centering
%         \includegraphics[width=\linewidth]{data/prestashop.png}
%         \caption{Prestashop}
%     \end{subfigure}
%     \hfill
%     \begin{subfigure}{0.32\textwidth}
%         \centering
%         \includegraphics[width=\linewidth]{data/joomla.png}
%         \caption{Joomla}
%     \end{subfigure}
%     \hfill
%     \begin{subfigure}{0.32\textwidth}
%         \centering
%         \includegraphics[width=\linewidth]{data/bagisto.png}
%         \caption{Bagisto}
%     \end{subfigure}

%     \caption{Code coverage comparison between standard grey-box (yellow), agnostic (red), and black-box (black) fuzzer after running for 24 hours. Due to the non-deterministic nature of fuzzing, each experiment runs three times.}
%     \label{fig:code-cov}
% \end{figure*}

\begin{figure*}[t]
    \centering
    \includegraphics[width=\linewidth]{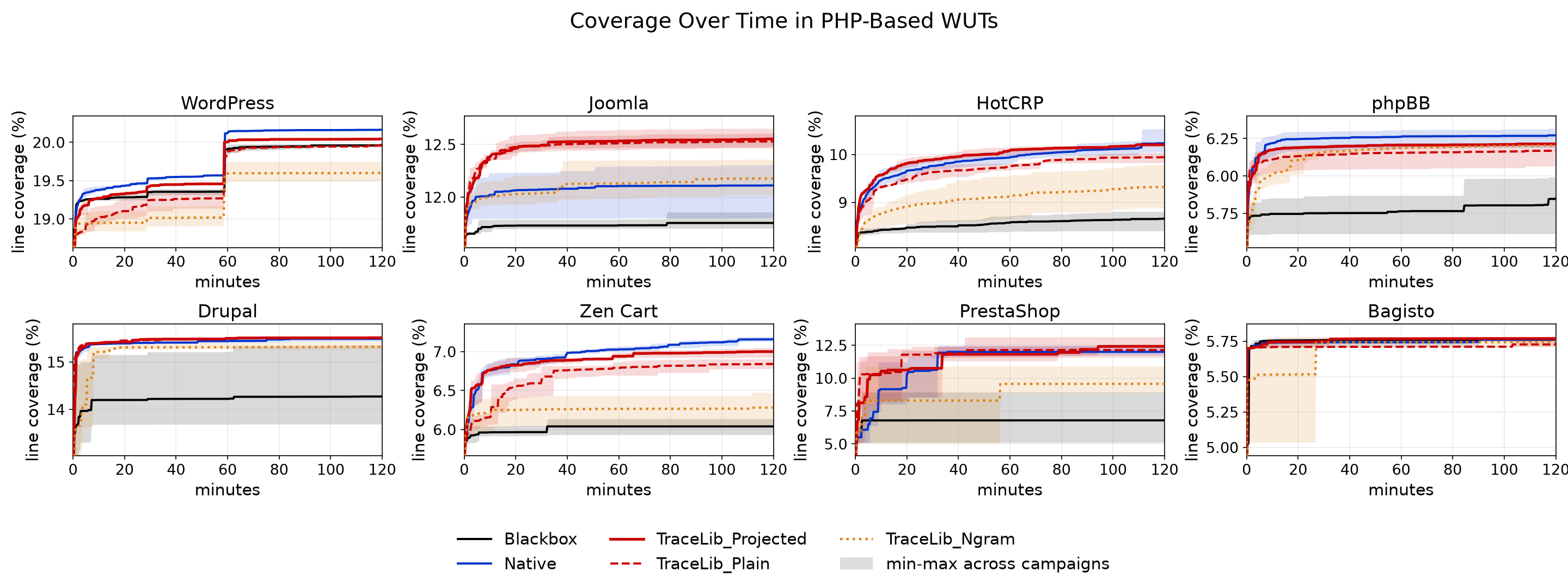}

    \caption{Coverage of the five feedback modes over the two-hour campaign in PHP-based WUTs. Since fuzzing results are non-deterministic, each campaign is run three times. While the Native leads on WordPress, HotCRP, phpBB, and ZenCart, the \tracelibprojected~leads on the other four.}
    \label{fig:code-cov}
\end{figure*}

\subsection{RQ1: Code Coverage}
\label{code-coverage-result}
In order to answer RQ1, we first prepare all of the PHP-based WUTs with WebFuzz’s code instrumentation to measure code coverage. Then, we launch the WebFuzz campaign and skip WebFuzz's crawling phase to remove endpoint discovery as a confound and to give the same start for each mode. Skipping the crawl phase is important because our initial experiments showed that some WUTs (e.g., PrestaShop) spent the entire time budget in the WebFuzz crawl phase and never started fuzzing.
Therefore, every feedback mode receives the same ten authenticated GET seeds, each validated with an HTTP 200 response, and has a 7,200-second wall-clock budget for the fuzzing campaign. This setting aligns with the binary fuzzing common practice, which also prepares several seeds for the fuzzing campaign.

\subsubsection{Overall Coverage}
\begin{figure}
    \centering
    \includegraphics[width=\linewidth]{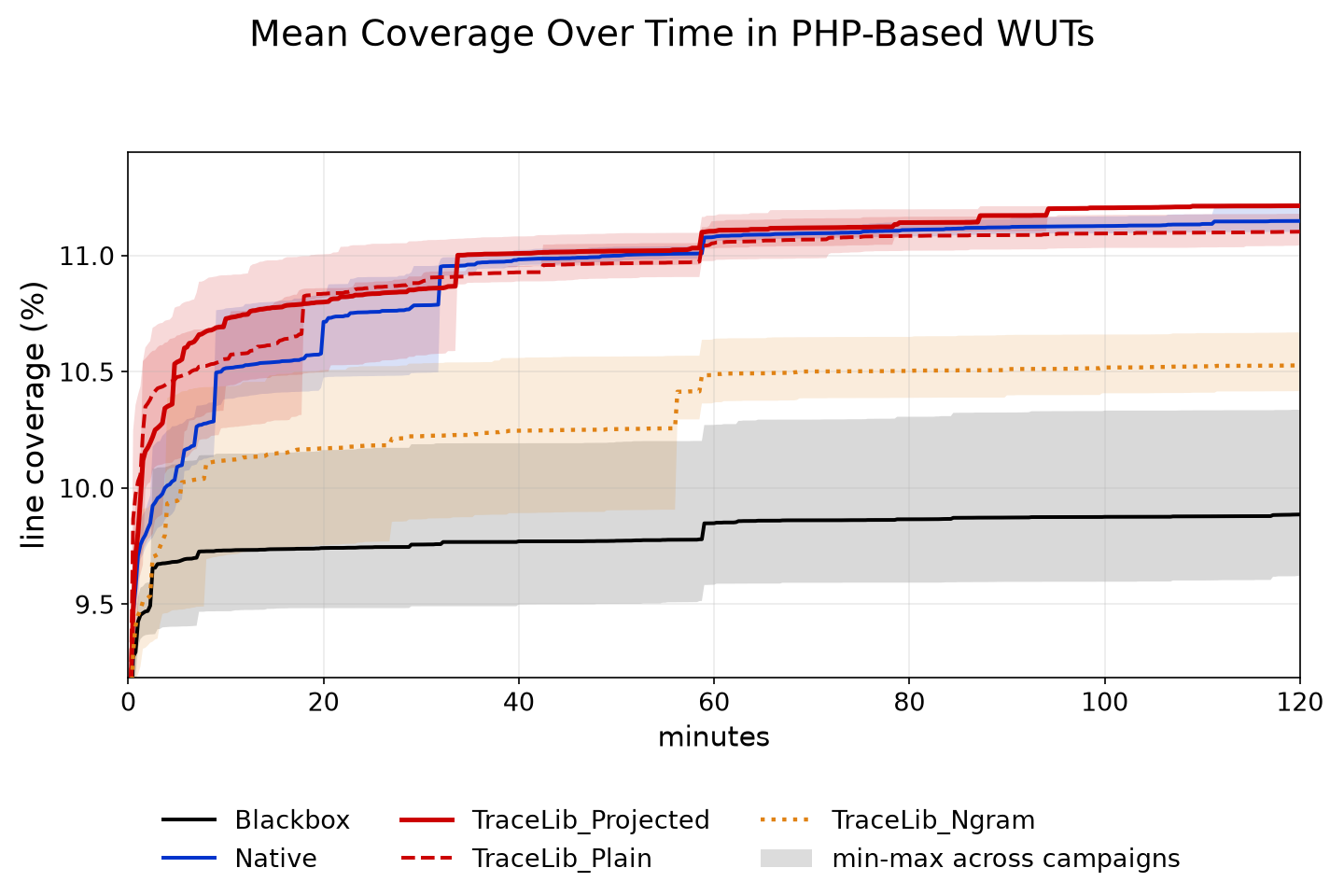}
    \caption{Averaged over the eight PHP-based WUTs and the three pooled campaigns, the code coverages achieved by both \tracelibprojected and \tracelibplain are similar to those from Native.}
    \label{fig:code-cov-ave}
\end{figure}

As shown in Figure \ref{fig:code-cov}, our proposed TraceLib (i.e., \tracelibplain~ and \tracelibprojected) and the Native achieve comparable coverage. Averaged over the eight PHP-based WUTs and the three campaigns (see Figure \ref{fig:code-cov-ave}), the Native attains 11.15\% while \tracelibprojected~attains 11.21\% and \tracelibplain~ 11.10\%. The difference between their averages is around 0.07 points, which is not significant, so they are tied. Per WUT, Native leads on four of the eight and \tracelibprojected~on the other four. These results show that our syscall-level feedback can match instrumentation-guided coverage while remaining portable, but the margin is within measurement noise, and the results do not establish that TraceLib dominates Native feedback.

% \begin{tcolorbox}[
%     colback=gray!10,
%     colframe=gray!40,
%     boxrule=0.5pt,
%     arc=2pt,
%     left=6pt,
%     right=6pt,
%     top=6pt,
%     bottom=6pt
% ]
% \centering
% \tracelib has comparable performance to \textit{Native} which can be summarized as follows (Figure~\ref{fig:code-cov}):\\
% \textbf{Key Takeway: TraceLib\_Projected $\approx$ Native $\approx$ TraceLib\_Plain
% $>$ TraceLib\_nGram $>$ Black-box}
% \end{tcolorbox}

% \subsubsection{Effect of Encoding Selectivity}
% The three TraceLib modes rank in the same order as their selectivity, and the order is stable across the WUTs. \tracelibprojected~ is at or above \tracelibplain~ on all eight WUTs, significantly so on HotCRP, PrestaShop and ZenCart, and above \tracelibngram~ on all eight, significantly so on six. This ordering is the inverse of how much each mode records, which is described in Section \ref{tracelib-modes}. \tracelibngram~ records every system call and more than one map position per event, and it retains between 3,731 and 25,979 inputs per WUT. \tracelibplain~ records every system call and one position per event, retaining between 62 and 448 inputs. \tracelibprojected records only argument-bearing calls and one position per event, retaining between 19 and 104 inputs. The encoding that records least therefore reaches the most coverage, and it does so from a corpus one to two orders of magnitude smaller than either comparator.

\subsubsection{\tracelibngram~Comparison}
\tracelibngram~is the representation adapted from the most recent work on system-call feedback for binary applications, and our measurements indicate that it does not transfer well to web workloads. It is the weakest of the three TraceLib modes in terms of coverage, and it is the only feedback source that performs worse than no feedback at all: on WordPress it reaches 19.60\% against black-box's 19.96\%.

The main reason is that its coverage representation is too sensitive for web activities. It records every system call the request performs and encodes the identity and the order of four or eight consecutive calls at a time, whereas \tracelibplain~encodes a transition between two consecutive calls and \tracelibprojected~encodes a transition between two calls that carry a relevant argument. A window of \tracelibngram~therefore summarizes a longer stretch of the syscall stream, which mostly contains calls that are unrelated to the application code.

\begin{figure}
    \centering
    \includegraphics[width=\linewidth]{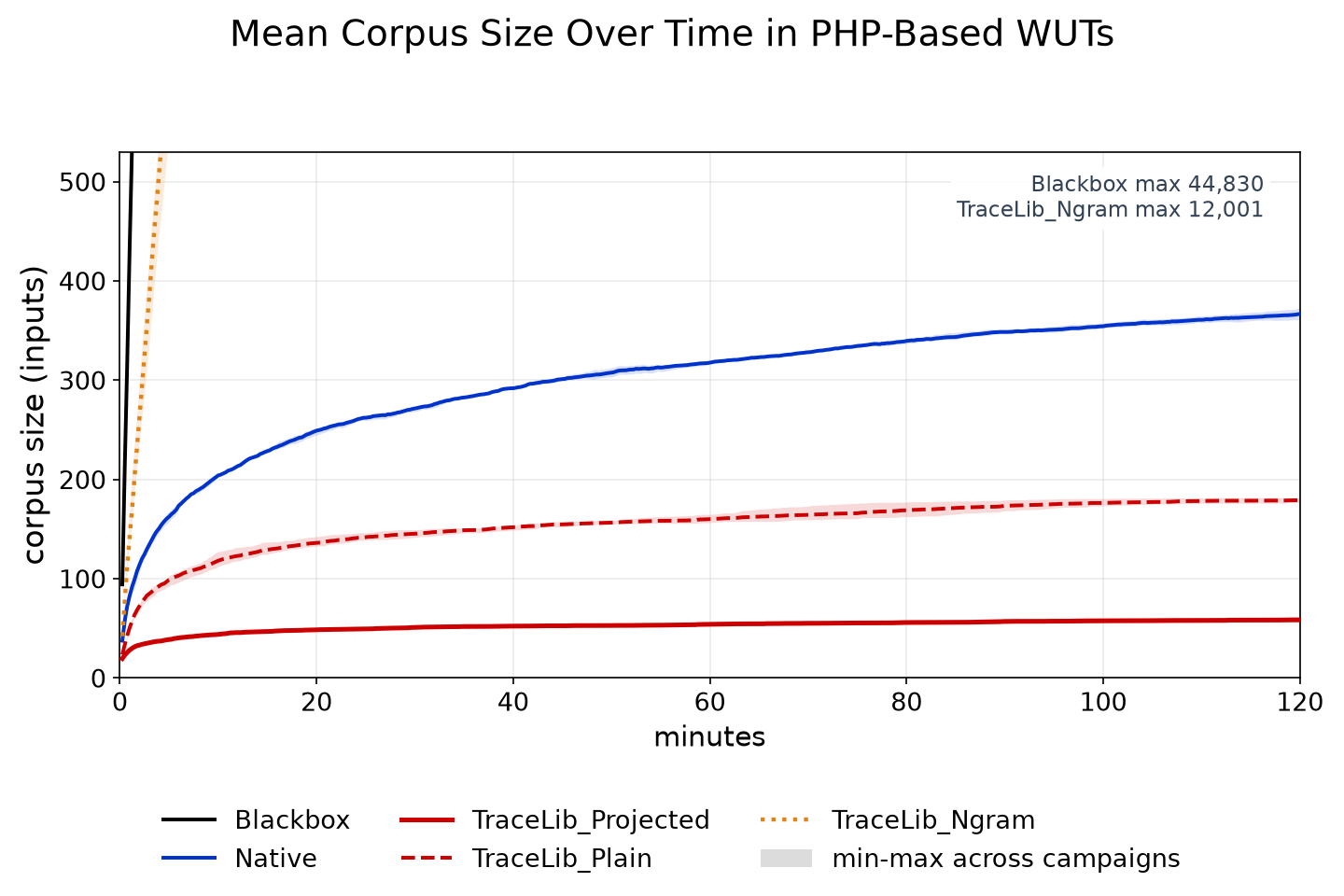}
    \caption{\tracelibprojected~keeps the smallest corpus while \tracelibngram~behaves similarly to the black-box mode.}
    \label{fig:corpus-ave}
\end{figure}

For example, a request in WordPress performs thousands of system calls, most of which reflect the state of the runtime or of the web environment rather than the request itself; \texttt{newfstatat} is a typical case. The number of these calls varies from one execution to the next even when the submitted requests are identical. Such variations are sufficient to change the window-based representation and, consequently, the coverage bitmap indices it produces. As a result, identical requests can produce different bitmaps, causing almost every pair of submitted requests to appear distinct to this mode. This leads to a much bigger yet less effective corpus, with behavior similar to the black-box configuration, which retains every request (see Figure \ref{fig:corpus-ave}).

\begin{table*}[htbp]
\centering
\caption{Pairwise classification results for \tracelibprojected, addressing RQ2 (Section \ref{tracelib-quality}). We deliberately configured the non-PHP applications to issue fewer HTTP requests because their platform-specific code-coverage tools crash when repeatedly measuring large numbers of requests.}
\label{tab:pairwise-results}
\resizebox{\textwidth}{!}{%
\begin{tabular}{l r r r r r r r r r r}
\toprule
\textbf{WUT} & \makecell{\textbf{Requests}\\\textbf{Completed}} & \makecell{\textbf{Request}\\\textbf{Pairs}} & \makecell{\textbf{True}\\\textbf{Positive}} & \makecell{\textbf{True}\\\textbf{Negative}} & \makecell{\textbf{False}\\\textbf{Positive}} & \makecell{\textbf{False}\\\textbf{Negative}} & \makecell{\textbf{TPR}\\\textbf{(\%)}} & \makecell{\textbf{FPR}\\\textbf{(\%)}} & \makecell{\textbf{TNR}\\\textbf{(\%)}} & \makecell{\textbf{FNR}\\\textbf{(\%)}} \\
\midrule
Zen Cart   & 1,000 & 499,500 & 475,909 & 6,575  & 5,766 & 11,250  & 97.69 & 46.72 & 53.28  & 2.31  \\
HotCRP     & 1,000 & 499,500 & 482,918 & 2,220  & 15    & 14,347  & 97.11 & 0.67  & 99.33  & 2.89  \\
Drupal     & 1,000 & 499,500 & 476,528 & 2,202  & 98    & 20,672  & 95.84 & 4.26  & 95.74  & 4.16  \\
PrestaShop & 1,000 & 499,500 & 449,880 & 23,346 & 800   & 25,474  & 94.64 & 3.31  & 96.69  & 5.36  \\
phpBB      & 1,000 & 499,500 & 470,184 & 1,656  & 18    & 27,642  & 94.45 & 1.08  & 98.92  & 5.55  \\
Joomla     & 810 & 327,645 & 287,265 & 14,492 & 174   & 25,714  & 91.78 & 1.19  & 98.81  & 8.22  \\
Bagisto    & 1,000 & 499,500 & 453,079 & 1,598  & 486   & 44,337  & 91.09 & 23.32 & 76.68  & 8.91  \\
WordPress  & 1,000 & 499,500 & 438,998 & 5,850  & 280   & 54,372  & 88.98 & 4.57  & 95.43  & 11.02 \\
% Ghost      & 50    & 1,225   & 1,077   & 85     & 3     & 60      & 94.72 & 3.41  & 96.59  & 5.28  \\
% Gitea      & 50    & 1,225   & 0       & 243    & 0     & 982     & 0.00  & 0.00  & 100.00 & 100.00 \\
% Casdoor    & 50    & 1,225   & 0       & 88     & 0     & 1,137   & 0.00  & 0.00  & 100.00 & 100.00 \\
% NodeBB     & 50    & 1,225   & 0       & 174    & 0     & 1,051   & 0.00  & 0.00  & 100.00 & 100.00 \\
% Redmine    & 49    & 1,128   & 702     & 51     & 3     & 372     & 65.36 & 5.56  & 94.44  & 34.64 \\
\midrule
Roller     & 50 & 1,225 & 1,111 & 22 & 86 & 6   & 99.5 & 79.6 & 20.4 & 0.5 \\
Gogs       & 50 & 1,225 & 1,181 & 7  & 27 & 10  & 99.2 & 79.4 & 20.6 & 0.8 \\
WikiJS     & 50 & 1,225 & 752   & 453 & 0 & 20  & 97.4 & 0.0  & 100.0 & 2.6 \\
Ghost      & 50 & 1,225 & 1,085 & 86 & 6  & 48  & 95.8 & 6.5  & 93.5 & 4.2  \\
Superset   & 50 & 1,225 & 1,070 & 79 & 16 & 60  & 94.7 & 16.8 & 83.2 & 5.3 \\
Huginn     & 50 & 1,225 & 1,111 & 30 & 0  & 84  & 93.0 & 0.0  & 100.0 & 7.0 \\
PetClinic  & 50 & 1,225 & 990   & 87 & 7  & 141 & 87.5 & 7.4  & 92.6 & 12.5 \\
Redmine    & 49 & 1,176 & 680   & 9  & 0  & 487 & 58.3 & 0.0  & 100.0 & 41.7 \\

\midrule
% \textbf{Total / Overall} & \textbf{8,249} & \textbf{3,830,173} & \textbf{3,536,540} & \textbf{58,580} & \textbf{7,643} & \textbf{227,410} & \textbf{93.96} & \textbf{11.54} & \textbf{88.46} & \textbf{6.04} \\

\textbf{Total / Overall} &
\textbf{8,209} &
\textbf{3,833,896} &
\textbf{3,542,741} &
\textbf{58,712} &
\textbf{7,779} &
\textbf{224,664} &
\textbf{94.04} &
\textbf{11.70} &
\textbf{88.30} &
\textbf{5.96} \\

\bottomrule
\end{tabular}%
}
\vspace{1pt}
\parbox{\textwidth}{\footnotesize
\textit{Note:} Request Pairs is calculated only from submitted HTTP requests that have both feedback (i.e., from TraceLib and Native), referred to as \texttt{requests completed} in the table. Consequently, Joomla uses only 810 complete requests, and Redmine uses only 49 complete requests, since some requests failed to produce TraceLib feedback.
%TPR/FNR are computed over actual positives ($TP+FN$); FPR/TNR over actual negatives ($TN+FP$).
}
\end{table*}

\subsection{RQ2: Per-Request Feedback Quality}
\label{tracelib-quality}
To evaluate TraceLib's effectiveness in accurately determining code coverage equivalence between web requests, we conducted a detailed comparative analysis against the coverage counters of all platforms (see the counter tools in Table \ref{tab:applications}). 
% the state-of-the-art coverage counters in PHP: the WebFuzz instrumentation and PCOV \cite{watkins_krakjoepcov_2025}.

\subsubsection{Methodology}
We collected a dataset of HTTP requests submitted by WebFuzz during the fuzzing phase. Each request entry includes the full URL, HTTP method, body parameters, the coverage results from TraceLib, and the coverage results from either WebFuzz's native instrumentation (for PHP) or third-party coverage counter tools (for non-PHP). 
%To gain these coverage results in the same experiments, the WUT is instrumented with WebFuzz, and then TraceLib and the coverage counter tools track the execution of the WUTs. 
To keep the coverage information concise and easier to process, each coverage result is hashed. The underlying principle is that if two requests trigger the same execution path (i.e., identical code coverage), the coverage counters should produce the same hash. Conversely, distinct coverage should yield different hashes. We compared all pairwise combinations of submitted requests and evaluated whether both methods agree on coverage equivalence.

\subsubsection{Positive Class Definition}
In our experiment, we primarily focus on identifying unique execution behaviors triggered by web requests. In fact, in coverage-guided fuzzing research, the goal is to discover new coverage, so the “\textbf{positive event}” is often defined as discovering uniqueness. Therefore, we define a true positive as occurring when both TraceLib and the ground truth (i.e., WebFuzz's AST coverage for PHP-based WUTs, or the platform-specific coverage tool for non-PHP WUTs) agree that two requests exercise different execution paths.

\begin{itemize}
    \item True Positive (TP): Both methods agree that two requests have different coverage.
    \item True Negative (TN): Both methods (i.e., TraceLib and the ground truth) agree that two requests have the same coverage.
    \item False Positive (FP): TraceLib reports different coverage while the ground truth reports the same coverage.
    \item False Negative (FN): TraceLib reports the same coverage while the ground truth does not.
\end{itemize}

% Therefore, we define the positive class as pairs of requests that exercise different execution paths (i.e., have different coverage).
% From these pairwise comparisons, we classified each result as:
% \begin{itemize}
%     \item True Positive (TP): Both methods (e.g., TraceLib and WebFuzz instrumentation) agree that two requests have the same coverage.
%     \item True Negative (TN): Both methods agree that two requests have different coverage.
%     \item False Positive (FP): TraceLib reports the same coverage while the other baseline (e.g., WebFuzz instrumentation) does not.
%     \item False Negative (FN): TraceLib reports different coverage while the other baseline reports the same coverage.
% \end{itemize}

% \subsubsection{WUT Configurations}
% To gain the coverage results from TraceLib, WebFuzz instrumentation, PCOV in the same experiments, the WUT is instrumented with WebFuzz, and then TraceLib and PCOV track the execution.

\subsubsection{Results}
As shown in Table \ref{tab:pairwise-results}, the evaluation revealed that our best proposed TraceLib (i.e., \tracelibprojected) achieved strong consistency with the ground-truth coverage across the dataset. Overall, \tracelibprojected~reached over 94\% True Positive Rates (TPR) with a quite low 11\% False Positive Rate (FPR) in coverage equivalence to the ground-truth coverage. This level of agreement indicates that \tracelib, despite operating at a lower abstraction level than code instrumentation, can capture behavior changes corresponding to meaningful execution differences. In other words, recording the transitions of system calls, executed files, and SQL queries enables language-agnostic deployment while preserving feedback accuracy.

\subsubsection{Small Negative Class}
Table \ref{tab:pairwise-results} also reveals an imbalance in our dataset, with substantially more positive pairs than negative pairs. A pair is labeled negative only when the coverage counter reports the same line coverage for both requests. Because two randomly selected requests are unlikely to produce exactly the same runtime coverage, most pairs are therefore labeled positive.

We do not rebalance the classes because this imbalance is a natural consequence of the fuzzing process. During fuzzing, newly generated requests are expected to explore new parts of the application and therefore produce new coverage. As a result, positive pairs naturally occur more frequently than negative pairs in our collected data.

% \subsubsection{False Positives on non-PHP WUTs}
% \label{sec:false-positive-non-php}
% TraceLib produces a false positive when it distinguishes two requests, while the corresponding coverage counter tool reports the same coverage. Consequently, the coverage tool is the primary oracle for this evaluation. For PHP-based WUTs, WebFuzz's native coverage instrumentation provides a stable reference by instrumenting code branches directly and capturing executed control-flow transitions. However, for non-PHP WUTs, WebFuzz does not support the same source-level instrumentation and instead relies on platform-specific coverage tools. These tools, such as \texttt{JaCoCo} and \texttt{c8}, primarily represent coverage as sets of executed lines, making them cannot distinguish behaviors that execute the same set of lines in a different order.

% As a result, a false positive in a non-PHP WUT does not necessarily indicate that the TraceLib feedback is incorrect. Two requests may execute the same set of lines while accessing different files or executing the same operations in a different order. \tracelibprojected~can distinguish these differences through system-call transitions even though the coverage tool reports them as identical. Consequently, even though Gogs and Roller have FPR more than 50\%, these false positives do not necessarily imply ineffective feedback because TraceLib captures behavioral distinctions that are not represented by the corresponding coverage oracles.

\subsubsection{False Positives on non-PHP WUTs} 
\label{sec:false-positives-non-php} 
Some non-PHP WUTs exhibit high FPRs, with Gogs and Roller having FPRs above 50\%. Recall that a false positive occurs when TraceLib distinguishes two requests while the corresponding coverage tool reports the same coverage. Because this metric treats the coverage tool as the ground truth, TraceLib is penalized for detecting fine-grained behavioral differences that the coverage tool cannot observe.

For PHP-based WUTs, WebFuzz's Edge Counter operates at code branches and records executed control-flow transitions, providing a relatively fine-grained oracle. For non-PHP WUTs, their coverage tools primarily represent coverage as a set of executed lines and do not preserve their execution order, as explained in Section \ref{sec:code-coverage-oracle}. Consequently, two requests may execute the same set of lines even though those lines are executed in a different order or a different number of times. TraceLib can distinguish such requests through system-call transitions, whereas the non-PHP coverage tools report them as identical. Therefore, the high FPRs for non-PHP WUTs stem from the coarser coverage oracle, not from ineffective feedback in TraceLib.

\subsection{RQ3: Time Overhead}
To answer RQ3, we set up five modes to evaluate the time overhead introduced during fuzzing. The black-box fuzzer executes the WUTs without any instrumentation or tracing, serving as a baseline for the fastest possible execution. We first record 1,000 requests generated by the black-box mode, together with their request sequence, and use these requests as a fixed replay workload for all other modes. This ensures that each mode executes the same inputs and allows us to isolate the overhead introduced by feedback collection. All TraceLib modes run the WUTs without code instrumentation but enable real-time observation of all system-call activities. Native, by contrast, executes instrumented WUTs that incorporate source-code instrumentation. 

\begin{figure*}
    \centering
    \includegraphics[width=\linewidth]{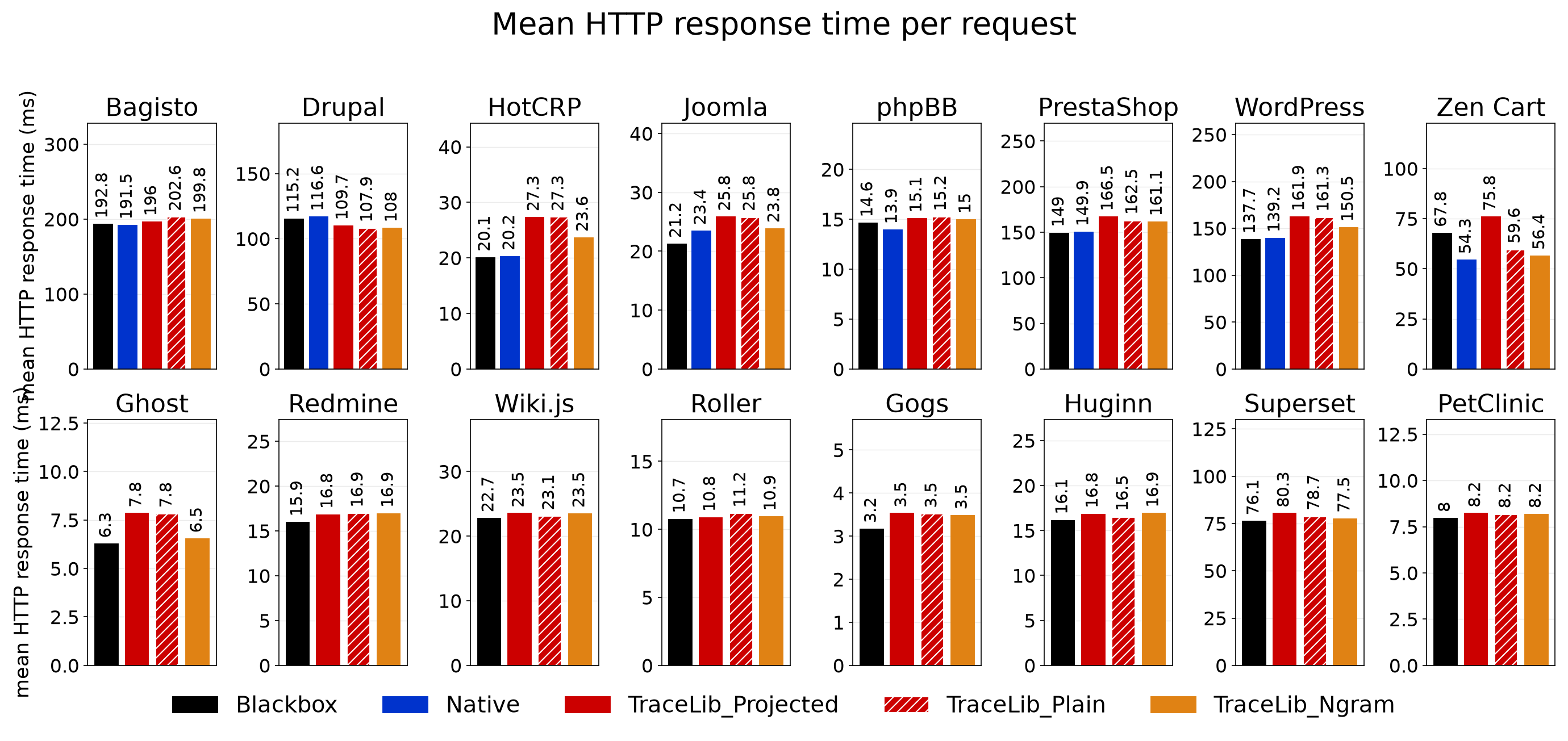}
    \caption{Mean per-request HTTP response time for each feedback mode over 1,000 byte-identical replayed requests. In general, system-call tracing adds approximately 0.81~ms of latency compared with the black-box mode.}
    \label{fig:response-time}
\end{figure*}

The result is shown in Figure \ref{fig:response-time}, which shows that the cost of observing a WUT through its system calls is small and largely independent of the approach applied to the resulting trace. Across the 48 TraceLib campaigns (16 WUTs × 3 modes), the mean server-side response time increases by approximately 0.81~ms, or 4.5\% of the corresponding black-box baseline. The overhead is concentrated in the most system-call-intensive PHP applications, where WordPress and PrestaShop incur up to 24.2 ms and 17.6 ms, respectively. These applications perform the largest amounts of file and database activity per request and therefore generate more system-call activity for the tracer to observe.

A small number of campaigns report a mean response time below the black-box baseline. This does not imply that code instrumentation or syscall tracing makes the WUT itself faster. Instead, the measurements are affected by variability in the server's runtime state and environment. This effect is particularly visible in applications with more complex execution environments, such as Drupal, phpBB, and Zen Cart, where factors such as cache state and connection-pool warm-up can affect server processing time between runs.

% As shown in Figure \ref{fig:response-time}, the \fuzzer with \tracelib increases the web processing time by approximately 5.7× compared to the instrumented WUTs. This overhead is expected. Unlike code instrumentation, which injects lightweight counters inside the WUT code, \tracelib operates at the system call level, requiring the operating system to intercept, record, and forward every relevant syscall event. Such OS-level tracing inherently introduces higher context-switching costs and larger monitoring overhead \cite{10.1145/3728874}.

% Despite this slowdown, the result aligns well with our expectations: syscall tracing is naturally more expensive than code instrumentation, yet it brings advantages. Most importantly, \tracelib eliminates the need for source code modification, language-specific build steps, or compiler support. As demonstrated in RQ1 and RQ2, this trade-off is beneficial. We can fuzz a web application immediately in its original form, regardless of the programming language or framework, without requiring any instrumentation pipeline, complex build systems, or tight integration with the application code. This greatly lowers the barrier to deploying feedback-guided fuzzing in diverse and realistic environments.

% Answering RQ3, we use those three fuzzers explained in Section \ref{code-coverage-result} to fuzzing WUTs. To measure the actual processing time, the black-box fuzzer tests the WUT without code instrumentation, which is supposed to run faster than the instrumented WUT. Agnostic fuzzer also tests the same WUT without instrumentation, but TraceLib monitoring is activated.

% As shown in Figure \ref{fig:processing-time}, AgnosticFuzz with TraceLib makes the web processing time around 5.7 times longer than the instrumented web applications. However, even though TraceLib slows down the WUTs processes, the benefit of requiring no instrumentation to start web fuzzing efficiently a web app is enormous, as answered in RQ1 and RQ2.

% \begin{figure}[t]
%     \centering
%     \includegraphics[width=1.0\linewidth]{data/time.png}
%     \caption{WUT's Processing Time}
%     \label{fig:processing-time}
% \end{figure}

\subsection{RQ4: Portability}
\begin{figure*}
    \centering
    \includegraphics[width=\linewidth]{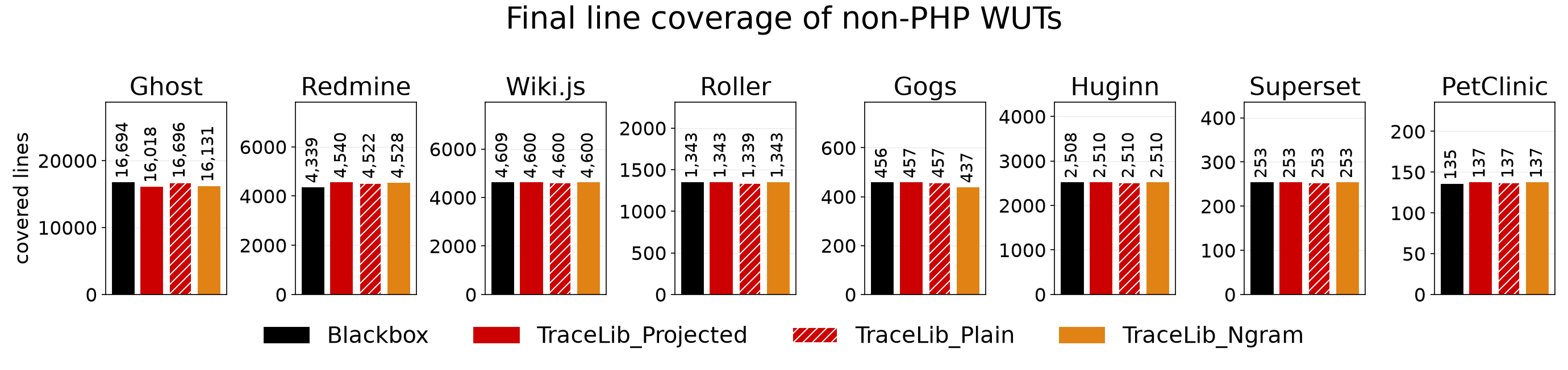}
    \caption{The final coverage of non-PHP WUTs. Coverage-over-time graphs are omitted because the per-language coverage counters crash under per-minute polling, so each counter is queried once at the end of the campaign.}
    \label{fig:nonphp-coverage}
\end{figure*}

Answering RQ4, Figure \ref{fig:nonphp-coverage} shows that our TraceLib feedback is practically useful across various runtimes, such as Node.js (Ghost and Wiki.js), Ruby (Redmine and Huginn), Java (Roller and PetClinic), Go (Gogs), and Python (Superset). In terms of final code coverage, our TraceLib slightly exceeds black-box on Redmine, PetClinic, Huginn, Ghost, and Gogs, ties it on Roller and Superset, and falls marginally behind on Wiki.js. These results establish portability and useful cross-runtime guidance, even though it is not an absolute advantage.
% TraceLib has demonstrated that it provides one consistent, source-independent feedback mechanism across multiple language ecosystems while substantially reducing corpus size.

This evaluation also shows that the coverage difference between TraceLib and black-box is smaller for the non-PHP WUTs than for the PHP-based WUTs. This difference is partly caused by the coverage oracles used for the two groups. 
As discussed in Section~\ref{sec:false-positives-non-php}, the coverage results from non-PHP coverage tools are coarser than the results from WebFuzz's Edge Counter in PHP. 
% As discussed in Section~\ref{sec:false-positives-non-php}, the non-PHP coverage tools used in our evaluation primarily represent the coverage as a set of code lines that are covered during execution and do not preserve the order in which those lines were executed. 
% % Consequently, different executions that traverse the same code lines in different orders can be reported as having identical coverage. TraceLib, in contrast, captures ordered system-call transitions and can distinguish some of these executions. 
% On the other hand, the WebFuzz's edge counter in PHP-based WUTs is able to record the set of code block transitions.
As a result, the non-PHP applications may appear to exhibit less behavioral variation, making the final coverage gap between TraceLib and black-box much smaller than in the PHP experiments. This should be interpreted partly as a limitation of the coverage reference.

% To this end, we selected a set of simple, open-source web applications implemented in Python (e.g., a Flask or Django app), Java (e.g., a Spring Boot microservice), and Node.js (e.g., an Express app). For the server infrastructure, we experimented with different process models, including a Python model using Gunicorn, a Java web server using embedded Tomcat, and a clustered Node.js process pool. In each case, we track the target application with TraceLib, launched the shared-memory feedback pipeline, and ran the agnostic fuzzer for a fixed time period (24 hours) using the same input seed corpus.

% Overall, these experiments confirm that 

% Answering RQ4, we runs the agnostic fuzzer in different settings, such as different web servers and different programming languages. We aim to show that the fuzzer is fully language and web server-agnostic. We highlight the effort needed to set up and execute the fuzzer. For that, we use open-source or simple web applications written in Python, Java, or Node.js.

\section{Limitations}
Although TraceLib achieves results comparable to Native mode, several limitations remain and should be addressed in future work.

\subsection{Single Active Request} 
A fundamental limitation is that TraceLib currently supports only one active HTTP request at a time, as explained in Section \ref{non-concurrent-request-req}. 
% This restriction does not mean that the WUT itself must be single-threaded; the WUT may still use Apache workers, PHP processes, Go goroutines, Node.js threads, or background tasks, and TraceLib keeps syscall state per target thread. Eventually, TraceLib combines those observations into one active request bitmap. 
This restriction is necessary for correct attribution because TraceLib aggregates all syscall observations into a single bitmap for the one active request. When a second request arrives while one is in progress, it replaces or finalizes the first request's record, and syscalls are then attributed to the wrong request. 
% If two external requests overlap, the second request can replace or finalize the first request’s control record, causing syscalls to be attributed to the wrong request. 
The cost of this design is that it prevents reliable multi-client throughput measurements and can miss concurrency-dependent bugs. A scalable design would require request- or connection-keyed contexts, separate bitmaps, and explicit handling for work that migrates across threads or continues asynchronously.

% \subsection{False Positive Rate}
% The second major limitation is the false-positive, or false-split, rate. Here, a false positive means that the Native oracle says two requests execute the same code, while TraceLib reports different bitmap coverage. In the alignment study (Section \ref{tracelib-quality}), TraceLib produced approximately 645,000 false splits but only 632 true negatives among code-equivalent pairs, giving roughly 99.9\% false-split rate for the same-code class. The system is therefore highly sensitive to differences, but poor at recognizing equivalent executions.

% These false splits have a direct fuzzing cost: requests are retained as “interesting” even though they do not extend application-code coverage, so the corpus grows, mutation effort is high, bitmap processing increases, and request throughput falls. Compared to TraceLib with N-Gram, our proposed TraceLib reduces this problem through shorter context and a compact corpus, but it does not eliminate it. TraceLib also risks observing syscalls generated by the Native oracle itself, such as coverage-map writes, which can make the reference instrumentation part of the feedback signal. 

\subsection{Dependence on Application Externalization}
\label{sec}
Our best approach, \tracelibprojected, records only system calls that carry either a monitored file path or a recognized SQL statement. Although \tracelibplain~records more system calls, this additional information does not necessarily provide better behavioral discrimination, because some application behavior is not exposed through system-call activity at all. For example, when a web application processes data that is already in memory, the computation may occur entirely in user space without crossing a system-call boundary. In such cases, neither TraceLib configuration can directly observe the corresponding behavior. In contrast, the Native mode uses AST-level instrumentation inserted at code branches, allowing it to observe the execution of those instrumented branches directly. 

Therefore, TraceLib's effectiveness depends partly on the WUT's characteristics, specifically how much of its control flow or state transitions are observable through system calls. This property varies substantially across our evaluated WUTs, which is reflected in the TPR of \tracelibprojected. The TPR ranges from approximately 58\% to 99\% across applications, indicating that TraceLib's ability depends on how much of the application's behavior is exposed through observable system calls. In particular, Redmine has the lowest TPR because it uses SQLite, where SQL queries are processed within the application rather than exposed as SQL statements in the system-call trace.
% and consequently leads to different levels of coverage compared with the Native mode.

% \subsection{Encrypted Data}
\subsection{Reduced Effectiveness on Encrypted Connections}
As explained before, \tracelibprojected~relies on observing SQL statements at the system-call boundary. Consequently, it cannot recover SQL text when the connection between the application and the database server is protected by Transport Layer Security (TLS), because the data crossing the network is encrypted before it reaches the system-call interface. In such cases, \tracelibprojected~cannot use SQL statements as feedback. Therefore, our proposed TraceLib is fully effective only when database information remains visible at the syscall boundary.

This limitation is particularly relevant because some WUTs use recent MySQL releases and drivers that may negotiate TLS, which can substantially reduce the amount of SQL information visible to TraceLib. In practice, this limitation can be mitigated in a fuzzing environment because the fuzzing setup is typically controlled by the web application developer. The developer can disable TLS for the database connection used during fuzzing, allowing SQL statements to cross the system-call boundary in plain text and making them available to \tracelibprojected.

\section{Related Work}
Coverage-guided fuzz testing has been extensively studied in the literature, traditionally focusing on binary applications. Compared to binary fuzzing, the application of grey-box fuzzing to web applications written in interpreted languages such as PHP and Python poses significantly more challenges related to instrumentation, state handling, and non-trivial input structures \cite{guler_atropos_2024, trickel_toss_2023, neef_what_2024}.

\subsection{Intrusive Modifications}
\textbf{WebFuzz} is a grey-box fuzzer that detects Cross-Site Scripting (XSS) vulnerabilities, but it requires changing the target’s PHP source code by inserting specific PHP code at the Abstract Syntax Tree (AST) level for coverage collection \cite{van_rooij_webfuzz_2021}. Modifying application code requires significant effort for testers and risks introducing unintended side effects \cite{neef_what_2024}.

\textbf{Witcher} generalizes grey-box fuzzing across multiple languages (e.g., PHP, Python, and Java), focusing on the detection of SQL and command injection vulnerabilities \cite{trickel_toss_2023}. While it avoids changing the application source code, it requires modifying the language’s interpreter (typically 1–5 lines of changes) for coverage data acquisition \cite{trickel_toss_2023}. Furthermore, Witcher uses intrusive techniques like binary-level hooking (e.g., hooking \texttt{libc}'s \texttt{recv} function for the database process or replacing \texttt{/bin/sh}) to escalate faults, requiring modifications to components outside the application code \cite{trickel_toss_2023}.

More recent grey-box systems demonstrate effectiveness while operating transparently to the application source code and the web server, confining instrumentation primarily to the interpreter. \textbf{Phuzz} is a modular, coverage-guided fuzzer for PHP web applications designed to find a wide array of server-side and client-side vulnerabilities \cite{neef_what_2024}. Phuzz requires no modification of the fuzzed application's source code, web server (such as Apache), database, or other external components. Instrumentation for coverage collection (via extensions like Xdebug or PCOV) and vulnerability detection (via function hooking and error monitoring) is handled entirely at the PHP interpreter level \cite{neef_what_2024}. \textbf{Atropos}, a snapshot-based, feedback-driven fuzzer for server-side PHP vulnerabilities, also localizes its advanced feedback mechanisms and bug oracles within the PHP interpreter to avoid application source modifications \cite{guler_atropos_2024}. Atropos uses hooks to infer necessary application-specific keys (e.g., variable names) and expected values (e.g., specific string tokens) at runtime. 
%For performance, Atropos utilizes fast snapshot restores via the NYX framework and communicates directly using the FastCGI interface, bypassing a traditional web server setup \cite{guler_atropos_2024}.

\subsection{Non-Intrusive Approaches}
Several recent approaches obtain fuzzing feedback without modifying the target application itself.
First, \textbf{SQLiFuzz} uses external proxies to guide SQL injection fuzzing \cite{10.1145/3808149}. It uses a database proxy to associate HTTP requests with SQL queries and to provide feedback for SQL injection testing. However, its observation focuses on the database layer and is limited to behavior reflected in SQL queries and database responses. In contrast, \tracelib observes system-call activity at the operating-system boundary, allowing it to capture not only database-related activity but also other application interactions, such as file access and network operations. This broader observation scope enables \tracelib to provide general-purpose feedback for web fuzzing rather than feedback specific to SQL injection. Thus, both approaches avoid modifying the application and its runtime, but \tracelib operates at a broader system level and can capture a wider range of application behavior than a database-only observation mechanism.

Next, Xiao et al. propose \textbf{system call pattern (SPC) coverage} as an alternative feedback mechanism for grey-box fuzzing~\cite{10.5555/3766078.3766399}. Designed primarily for binary-only COTS applications, their approach uses hashed $N$-gram patterns of system calls and their parameters. This work is closely related to \tracelib because both use system-call behavior as fuzzing feedback. However, \textbf{SPC} was designed around whole-program binary execution, whereas \tracelib must associate system calls with individual HTTP requests in a long-running server and selectively retain argument information that is meaningful for web applications. Therefore, we include an N-gram configuration, \textbf{\tracelibngram}, adapted from \textbf{SPC} as a comparator in both our system design and evaluation.

\subsection{Specialized Formats}
There are also fuzzers that focus on specific Web application formats or domains. \textbf{BackREST} \cite{gauthier_experience_2022} and \textbf{RESTler} \cite{atlidakis_restler_2019} are model-based grey-box and black-box fuzzers, respectively, designed primarily for REST APIs, often requiring OpenAPI specifications. They are similar to \textbf{Restest}\cite{martin-lopez_restest_2021}, \textbf{RestTestGen} \cite{viglianisi_resttestgen_2020}, \textbf{Nautilus} \cite{deng_nautilus_2023}, and \textbf{Schemathesis} \cite{hatfield-dodds_deriving_2022} in the sense of focusing primarily on OpenAPI specifications. These tools differ from full application fuzzers such as ours, Phuzz and Atropos, which target the interpretation layer of complex, dynamically structured web pages. 

\section{Conclusion}
This paper introduced a language-agnostic web fuzzing approach based on syscall-level tracing, addressing the limitations of traditional fuzzers that rely on language-specific instrumentation. Through TraceLib, our approach captures low-level execution behavior and enables web fuzzers to leverage feedback from the internal execution of web applications across diverse web stacks, without requiring modifications to application code or runtime environments. Our evaluation across 8 popular PHP-based WUTs and 8 non-PHP WUTs shows that the proposed \tracelibprojected achieves code coverage and corpus growth comparable to the Native grey-box fuzzers, while substantially outperforming the black-box baseline on PHP-based WUTs. In some cases, it even surpasses language-instrumented fuzzing, demonstrating the potential of syscall-level feedback to expose more nuanced execution variations. Moreover, the minimal setup effort required highlights the replicability and usability of our approach in real-world settings.

Overall, our results suggest that syscall-level tracing is a promising foundation for scalable, portable, and effective web fuzzing. Future work includes addressing the non-concurrent request requirement and the dependence on application externalization.

% \section*{Acknowledgments}
% We acknowledge the contributions of the bachelor’s and master’s students involved in the design and implementation phases of this work.

\section*{Acknowledgements}
The first author has received scholarship funding from the Center for Higher Education Funding and Assessment (PPAPT) and the Indonesia Endowment Fund for Education (LPDP) under the Indonesian Education Scholarship (BPI) schema.

\section*{Data Availability}
To promote transparency and reproducibility, we follow the Open Science policies.
\begin{enumerate}
    \item \textbf{\textit{Artefacts}}: The implementation of TraceLib is available in our public repository\footnote{\url{https://github.com/websecfuzz/tracelib}}. The repository also contains scripts for setting up the evaluation environment.
    % \item \textbf{\textit{Real-world applications}}: We evaluate AgnosticFuzzer on publicly available applications, namely WordPress, Drupal, Zencart, Prestashop, Joomla, and Bagisto. Instructions and setup scripts are included in the artefact package.
    \item \textbf{\textit{Reproduction package}}: A Docker-based package for web environments is included to facilitate an easier replication process.
\end{enumerate}

\section*{AI Usage}
During the preparation of this work, the authors used Anthropic Claude Code for code generation, debugging, refinement, and draft brainstorming. All final analysis, validations, and intellectual contributions were made by the authors. 

% \bibliographystyle{plain}
% \bibliography{data/references} % refers to references.bib
% \bibliographystyle{ACM-Reference-Format}
% \bibliography{data/references}

\bibliographystyle{cas-model2-names}
% \bibliography{cas-refs}
\bibliography{data/references}

% \appendices

% \section{Data Availability}
% \lipsum[1-4]
% \section{Theorem proofs}
% \lipsum[5-6]

% \vspace{12pt}
% \color{red}
% IEEE conference templates contain guidance text for composing and formatting conference papers. Please ensure that all template text is removed from your conference paper prior to submission to the conference. Failure to remove the template text from your paper may result in your paper not being published.

\end{document}